\documentclass[aps,floatfix,showpacs,amsmath,amssymb]{revtex4}
\usepackage{natbib}
\usepackage{dcolumn}
\usepackage{graphicx}
\usepackage{color}
\usepackage[usenames,dvipsnames]{xcolor}
\usepackage[normalem]{ulem}
\usepackage{multirow}
\usepackage{ulem}
\usepackage{enumitem}

\newcommand{\apjs}{Astrophys.~J.~Supp.}
\newcommand{\be}{\begin{equation}}
\newcommand{\ee}{\end{equation}}
\newcommand{\bea}{\begin{eqnarray}}
\newcommand{\eea}{\end{eqnarray}}

\def\[{\begin{equation}}
\def\]{\end{equation}}
\begin{document}
%
\title{Large-scale growth evolution in the Szekeres inhomogeneous cosmological models with comparison to growth data}
\author{Austin Peel\footnote{austin.peel@utdallas.edu}}
\author{Mustapha Ishak\footnote{mishak@utdallas.edu}}
\author{M. A. Troxel\footnote{troxel@utdallas.edu}}
\affiliation{
Department of Physics, The University of Texas at Dallas, Richardson, Texas 75083, USA}
\date{\today}
%
\begin{abstract}
We use the Szekeres inhomogeneous cosmological models to study the growth of large-scale structure in the universe including nonzero spatial curvature and a cosmological constant. In particular, we use the Goode and Wainwright formulation of the solution, as in this form the models can be considered to represent exact nonlinear perturbations of an averaged background. We identify a density contrast in both classes I and II of the models, for which we derive growth evolution equations. By including $\Lambda$, the time evolution of the density contrast as well as kinematic quantities of interest can be tracked through the matter- and $\Lambda$-dominated cosmic eras up to the present and into the future. In class I, we consider a localized cosmic structure representing an overdensity neighboring a central void, surrounded by an almost Friedmann-Lema\^itre-Robertson-Walker background, while for class II, the exact perturbations exist globally. In various models of class I and class II, the growth rate is found to be stronger in the matter-dominated era than that of the standard lambda-cold dark matter ($\Lambda$CDM) cosmology, and it is suppressed at later times due to the presence of the cosmological constant. We find that there are Szekeres models able to provide a growth history similar to that of $\Lambda$CDM while requiring less matter content and nonzero spatial curvature, which speaks to the importance of including the effects of large-scale inhomogeneities in analyzing the growth of large-scale structure. Using data for the growth factor $f$ from redshift space distortions and the Lyman-$\alpha$ forest, we obtain best fit parameters for class II models and compare their ability to match observations with $\Lambda$CDM. We find that there is negligible difference between best fit Szekeres models with no priors and those for $\Lambda$CDM, both including and excluding Lyman-$\alpha$ data. We also find that the standard growth index $\gamma$ parametrization cannot be applied in a simple way to the growth in Szekeres models, so a direct comparison of the function $f$ to the data is performed. We conclude that the Szekeres models can provide an exact framework for the analysis of large-scale growth data that includes inhomogeneities and allows for different interpretations of observations.
\end{abstract} 
\pacs{98.80.Es,98.80.-k,95.30.Sf}
\maketitle
%
\section{Introduction}\label{intro}
%
Studying inhomogeneous cosmological models is becoming increasingly important as the available cosmological data expand and improve. These exact solutions to Einstein's field equations provide frameworks to analyze the data with a wider range of possible interpretations. They can also be used to represent nonlinearities, which cannot be described using a Friedmann-Lema\^itre-Robertson-Walker (FLRW) model plus linear perturbations. The real universe is lumpy and exhibits a striking variety of small- and large-scale inhomogeneous structures, such as large voids, clusters, and superclusters of galaxies. Some of these superstructures can be as large as 5\% or 10\% of the Hubble scale, including the well-known Pisces-Cetus Supercluster Complex and the Sloan Digital Sky Survey Great Wall.

Exact solutions to Einstein's equations representing inhomogeneous cosmological models have been the subject of a number of theoretical studies \cite{SKMHH2003,Krasinski1997}; however, their comparison to observational data is only a recently emerging field. For example, although spherically symmetric, the Lema\^itre-Tolman-Bondi (LTB) solutions have achieved some proof-of-principle successes in their application to cosmology, where certain models have been shown to be able to fit type Ia supernova and baryon acoustic oscillations (BAO) data, as well as the cosmic microwave background power spectrum \cite{Iguchietal2002,Alnesetal2006,Enqvist&Mattsson2007,Garfinkle2006,Kaietal2007,Biswasetal2007,Tanimoto&Nambu2007,Enqvist2008,Chung&Romano2006,Garciaetal2008,Biswasetal2010,Quartin&Amendola2010,Celerier2007}. 

We use in this work the Szekeres inhomogeneous cosmological models. These exact solutions have no Killing vectors (i.e., no symmetries) 
and are well suited to represent the lumpy universe we observe. They have a broad potential in cosmology to provide a more realistic description of the Universe \cite{Bonnoretal1977,Ellis&VanElst1998}. The solution was first derived by Szekeres \cite{Szekeres1,Szekeres2} from a general metric with irrotational dust as the source of the spacetime. Several authors have studied the models analytically and numerically; a partial list includes 
\cite{Bonnor&Tomimura1976,Bonnoretal1977,Bonnor1986,Sussman&Triginer1999,Mena&Tavakol1999,Hellaby&Krasinski2006,Hellaby&Krasinski2002,Bolejko2006,Bolejko2007,Nolan&Debnath2007,Ishaketal2008,Nwankwoetal2011,Meures&Bruni2011,Goode&Wainwright1982a,Goode&Wainwright1982b,SussmanBolejko,BolejkoSussman,Ishak&Peel2012}. The Szekeres models have been extended to include pressure in \cite{Szafron1977} but with some known limitations on the state variables, and they were also extended by \cite{Barrow&SS1984} to include a cosmological constant.

The expansion and growth histories must both be consistent with observational data in any cosmological model that purports to describe the universe. Cosmological distances are necessary to understand the expansion history, and distances versus redshift have been previously studied for Szekeres models \cite{Ishaketal2008,Bolejko&Celerier2010,Nwankwoetal2011}. The growth rate of large-scale structure in the matter-dominated era for flat and curved cases of class I and class II Szekeres models have been studied in our earlier paper \cite{Ishak&Peel2012}. In that work, a formulation of the models due to Goode and Wainwright (GW) was used \cite{Goode&Wainwright1982a,Goode&Wainwright1982b}, which is well suited to studying structure growth in the universe. In the GW representation, Szekeres models can be interpreted as exact nonlinear perturbations of some smooth associated FLRW background. Recently, the authors of \cite{Meures&Bruni2011} extended the work of GW for class II flat models with a cosmological constant and examined nonlinear inhomogeneities in a $\Lambda$CDM background. Specializing to the spatially flat case, they showed that the inclusion of a cosmological constant allows models in some cases to avoid shell-crossing singularities. The evolution of structure in Szekeres models has also been studied via invariant density contrast indicators \cite{Mena&Tavakol1999,Bolejko2006,Bolejko2007}, but we take a different approach here.

The present paper generalizes our previous work \cite{Ishak&Peel2012} to include a cosmological constant in exploring the growth history of Szekeres models. Thus, the analysis now applies fully to the time since recombination up to the present and into the future, both  dominated by the cosmological constant. Since we use an exact framework instead of first-order linear perturbations, as is done with FLRW models, our work is not limited to the linear regime or only first-order terms in the density contrast. We derive differential equations for the density contrast in both classes (converting to the growth rate variable $G$ in class II), as well as expressions for the shear, expansion, and tidal gravitational field in terms of the density contrast and practical measurable cosmological parameters. In this form, the equations provide a more straightforward connection between theory and observation than do the metric functions on their own. We treat the class I formalism in much the same way as in \cite{Ishak&Peel2012} by introducing a quasilocal average density \cite{Sussman1,SussmanAIP,Sussman2,SussmanBolejko} in order to define a density contrast. To compare Szekeres growth to observations, we use data for the growth factor $f$ from redshift space distortions and the Lyman-$\alpha$ forest to obtain best fit parameters for class II models and compare them to those of the concordance $\Lambda$CDM cosmology. We also explore the applicability of the usual growth index $\gamma$ to parametrize the growth factor in Szekeres. 

The outline of the paper is as follows. We first present in Sec. \ref{GWform} the GW formulation of the Szekeres solution. We then derive the exact growth equations for class I and class II models including a cosmological constant in Sec. \ref{growthEqns}. Results and discussion from integrating the equations are given in Sec. \ref{growthPlots}, and in Sec. \ref{scalars} we examine the early- and late-time behaviors of scalar quantities that appear in the Raychaudhuri equation and can be identified as the cause of the strengthened Szekeres growth. We derive the equation for the growth factor in class II models in Sec. \ref{growthFactor} and compare it to cosmological data along with the standard $\Lambda$CDM cosmology. We conclude in Sec. \ref{conclusion}. Units are chosen so that $8\pi G = c = 1$ throughout the paper.
%
\section{The Szekeres Cosmological Models}\label{GWform}
%
We begin with an introduction of the Szekeres exact solution to Einstein's field equations in the form due to Goode and Wainwright (GW) \cite{Goode&Wainwright1982a,Goode&Wainwright1982b} and with a cosmological constant \cite{Meures&Bruni2011}. The GW form is a reformulation of the original solution discovered by Szekeres \cite{Szekeres1,Szekeres2}. A third formulation, discussed briefly in Appendix B, is also common in the literature and has a form similar to that of the LTB metric. It has been used in, for example, \cite{Hellaby&Krasinski2002,Bolejko2006,Plebanski&Krasinski2006,Bolejko2007,Ishaketal2008,Bolejko&Celerier2010,BolejkoSussman,Nwankwoetal2011}. The GW formulation is well suited for studying structure growth, since, as was remarked in \cite{Goode&Wainwright1982b}, one can consider the models as nonlinear exact perturbations of some averaged background. This background can be associated with a corresponding unperturbed FLRW model. The evolution equations in the Szekeres models can then be compared to linear perturbations of this associated FLRW model. It is worth noting that this association between the Szekeres background and an FLRW model is not the same as the case where a given inhomogeneous model becomes an FLRW model when some of the metric parameters (or certain coordinates) are taken to some large limit. One has to make further explicit definitions of some of the metric functions in order to implement such limits if needed. The full GW formulation (without a cosmological constant) can be found in the original papers \cite{Goode&Wainwright1982a,Goode&Wainwright1982b}, and we therefore present only a brief introduction in order to be self-contained and to set the notation. The GW form of the Szekeres metric is
\be
	ds^2= -\mathrm{d}t^2+S^2\left[e^{2\nu}(\mathrm{d}\tilde{x}^2+\mathrm{d}\tilde{y}^2)+H^2W^2 \mathrm{d}r^2\right],
\label{eq:metric}
\ee
where it is assumed that the functions $S$, $H$, and $W$ are all positive. The coordinates of the metric are comoving and synchronous, so the cosmic dust fluid has four-velocity components $u^\alpha=\delta^\alpha_0$. 

We note that the coordinates $\tilde{x}$ and $\tilde{y}$ are not the ``Cartesian'' coordinates $x$ and $y$. We have added a tilde on the GW coordinates for the Szekeres models in order to make this distinction clear to the reader. The GW coordinates $\tilde{x}$ and $\tilde{y}$ are the result of stereographic projections, and their transformation relations to known coordinate systems are given in Appendix B, see also  \cite{Plebanski&Krasinski2006}. Next, for clarity we labeled the third GW spacelike coordinate as $r$ instead of the $z$ that was used in previous papers \cite{Goode&Wainwright1982a,Goode&Wainwright1982b,Ishak&Peel2012}. 

From the outset, we will refer to the scaling function of the spatial part of the metric as $a$ instead of $S$ to facilitate comparisons with $\Lambda$CDM. The metric function $H$ depends on all four coordinates and can be written as the difference of two functions, only one of which carries the time dependence,
\be
	H(t,r,\tilde{x},\tilde{y})=A(r,\tilde{x},\tilde{y})-F(t,r).
\label{eq:H}
\ee
The function $F$ satisfies the second-order linear differential equation
\be
	\ddot{F}+ 2\frac{\dot{a}}{a}\dot{F}-\frac{3M}{a^3}F=0,
\label{Ray1}
\ee
where $\dot{}\equiv\partial/\partial t$ and $M(r)$ is an arbitrary but sufficiently smooth function. This equation can be derived from either the field equations or from the Raychaudhuri equation for irrotational dust \cite{Goode&Wainwright1982b,Raychaudhuri1955}. There are two linearly independent solutions to Eq. (\ref{Ray1}) denoted $f_+$ and $f_-$ (the so-called growing and decaying modes, respectively), and so $F$ can be written generally as
\be
	F=\beta_+f_+ + \beta_-f_-,
\ee
where $\beta_{\pm}$ are functions of $r$.

The models divide naturally into two classes according to the metric function dependencies, and the exact forms of the functions $A$, $\nu$, $f_{\pm}$, $\beta_{\pm}$, etc. for each class are given in Appendix A. In class I, which is the more general of the two, $a=a(t,r)$, $f_{\pm}=f_{\pm}(t,r)$, $M=M(r)$, and $W=W(r)$. In class II, these functions lose their $r$ dependence so that $a=a(t)$, $f_\pm=f_\pm(t)$, $M=$ const, and $W=1$.
 
The scale function obeys a generalized Friedmann equation
\be
	\frac{\dot{a}^2}{a^2}=\frac{2M}{a^3}+\frac{\Lambda}{3}-\frac{k}{a^2},
\label{Friedmann}
\ee
where $k=0,\pm 1$. Equations (\ref{Ray1}) and (\ref{Friedmann}) govern the time evolution of the models and apply generally to both classes. By Eq. (\ref{Friedmann}), we see that the class I $a(t,r)$ satisfies the usual Friedmann equation of FLRW for every value of $r$ so that the surfaces of constant $r$ evolve independently in time.

The matter density in both classes is given by
\be
	\rho=\frac{6MA}{a^3H}=\frac{6M}{a^3}\left(1+\frac{F}{H} \right).
\label{density}
\ee
In class I, we can identify an exact density contrast $\hat\delta=F/H$, which measures deviations from a background density. However, the interpretation of this $\hat\delta$ is different from that of the usual FLRW $\delta=(\rho-\bar{\rho})/\bar{\rho}$, where $\bar{\rho}$ is the average density of the space. $\hat\delta$ at some event $(t_0,r_0,\tilde{x}_0,\tilde{y}_0)$ compares the density there to the average density inside the surface defined by $t_0$ and $r_0$, instead of to some overall background or limiting value of $\rho$ where the solution becomes FLRW. Therefore, in order to facilitate the comparison between the growth in Szekeres class I and the perturbed FLRW, we use a Szekeres model as specified in Appendix C, representing a large-scale cosmic  structure surrounded by an FLRW model. On the other hand, in class II, $a$ and $M$ have no $r$ dependence, so it is natural to define a background density by $\bar{\rho}(t)=6M/a^3(t)$. By keeping $\delta=F/H$ (no hat to distinguish it from class I), the density contrast then describes deviations from some smooth underlying background like it does in FLRW. But while $\delta$ is written in the same way as in FLRW, 
a spatial profile is still needed in the Szekeres class II in order to facilitate the comparison to the FLRW plus perturbations scheme. This can be done for class II using the metric functions $\beta_{\pm}(r)$. Indeed, one recalls that both classes of Szekeres solution become FLRW when the functions $\beta_{\pm}(r)$ are zero. This is in fact the necessary and sufficient condition \cite{Goode&Wainwright1982b}. It is in this case, that $F$ and $\delta$ become identically zero, and the density assumes a similar form, $\rho=6M/a^3$, to that of FLRW. Since class I has been claimed to be more relevant than class II for astrophysical applications \cite{Plebanski&Krasinski2006}, we chose to implement the profile modeling in class I only and use the function $M(r)$, while for class II, we integrate the growth over time for a fixed value of $r$ with no further modeling of the functions $\beta_{\pm}(r)$. Finally, we note that the density is constrained to have the same sign as $M$, and so to have a physically reasonable solution we restrict our investigation in this work to models with positive $M$. 

\section{Cosmological Evolution of Large-Scale Growth using the Szekeres Models}\label{growthEqns}
%
In class II, the density contrast offers a straightforward comparison between exact Szekeres models and the linearly perturbed FLRW equation in $\delta$. However, we first address large-scale structure growth in class I via $\hat\delta$, as it is the more general of the two classes and requires more care. The treatment here is similar to our previous work \cite{Ishak&Peel2012} and employs, for class I, the quasilocal variables introduced and used in \cite{Sussman1,SussmanAIP,Sussman2,SussmanBolejko}. The main difference between this and our previous work is that whereas before we considered only the matter-dominated era, we now derive growth equations that include a cosmological constant, and so our analyses apply from the time of last scattering up to the $\Lambda$-dominated present and future evolutions. Finally, we recall that the Szekeres class I flat models have no growing modes \cite{Goode&Wainwright1982a,Goode&Wainwright1982b}, so we left them out. 
%
\subsection{Growth equations in curved Szekeres class I models with a cosmological constant}\label{growthEqnsClassI}
%
In order to treat the two Szekeres classes similarly, we make some definitions for class I so that the matter density may be split into a background and a density contrast, as mentioned at the end of Sec. \ref{GWform}. We introduce the quasilocal average density \cite{Sussman1,SussmanAIP,Sussman2,SussmanBolejko}
\be\
	\rho_q(t,r)= \frac{\int_{\tilde{y}}\int_{\tilde{x}}\int_r \mathcal{F} \rho(t,r,\tilde{x},\tilde{y})\sqrt{-h}\, \mathrm{d}r\, \mathrm{d}\tilde{x}\, \mathrm{d}\tilde{y}}{\int_{\tilde{y}}\int_{\tilde{x}}\int_r \mathcal{F} \sqrt{-h}\,\mathrm{d}r\, \mathrm{d}\tilde{x} \,\mathrm{d}\tilde{y}}=\langle \rho \rangle _{q\,\mathcal{D}[r]}(t)
\label{quasi-local-density}
\ee
over the domain $\mathcal{D}[r]$, which is bounded by an $r=$ const surface and is a subset of the three-dimensional hypersurfaces of constant $t$. The function $\mathcal{F}$ is a weighting factor, and the projection tensor ($h_{\mu\nu}=g_{\mu\nu} +u_\mu u_\nu$) has determinant $h$ of its three-dimensional part. The density contrast $\hat{\delta}$ can then be defined as usual but with $\rho_{q}(t,r)$ serving as the background average,
\be
	\hat\delta (t,r,\tilde{x},\tilde{y})\equiv \frac{\rho(t,r,\tilde{x},\tilde{y})\,\, - \rho_{q}(t,r)}{{\rho}_{q}(t,r)}.
\label{delta_hat}
\ee
As discussed in \cite{Sussman1,SussmanAIP,Sussman2,SussmanBolejko}, $\rho_q$ is a coordinate invariant quantity that can be shown to be given by
\be
	\rho_q(t,r)=\frac{6M(r)}{a^3(t,r)}.
\label{Szekeres_Rho_q}
\ee
By Eq. (\ref{delta_hat}), we then have 
\be
	\rho(t,r,\tilde{x},\tilde{y})= \rho_q(t,r)[1+\hat\delta(t,r,\tilde{x},\tilde{y})],
\ee 
which, upon comparing to Eq. (\ref{density}), allows us to make the identification
\be
	\hat\delta(t,r,\tilde{x},\tilde{y}) =\frac{F(t,r)}{H(t,r,\tilde{x},\tilde{y})}.
\label{identification2}
\ee
Since $\rho_q$ is a coordinate invariant quantity, so too must $\hat\delta$ be. As was done in \cite{Ishak&Peel2012}, the time differential equation for $F$ [Eq. (\ref{Ray1})] can be recast in terms of $\hat\delta$ to arrive at the class I evolution equation for the density contrast,
\be
	\ddot{\hat\delta}+2\frac{\dot{a}(t,r)}{a(t,r)}\dot{\hat\delta}-\frac{3M(r)}{{a^3(t,r)}}\hat\delta-\frac{2}{1+\hat\delta}{{\dot{\hat\delta}^2}}-3\frac{3M(r)}{a^3(t,r)}\hat\delta^2=0.
\label{growth1_classI}
\ee

The form of the above equation is suggestive of linear density perturbations in FLRW. To clarify the connection, we use Eq. (\ref{Friedmann}) to define generalized density parameters analogous to those of $\Lambda$CDM. Such a generalization is similar to what has been done in previous works for the LTB inhomogeneous models in order to compare them to observations (e.g., see \cite{Enqvist&Mattsson2007,Biswasetal2010,Quartin&Amendola2010,Celerier2007} and references therein). Equation (\ref{Friedmann}), with its explicit coordinate dependencies, in class I is
\be
	\frac{\dot{a}^2(t,r)}{a^2(t,r)} \equiv \mathbb{H}^2(t,r) = \frac{2M(r)}{a^3(t,r)} +\frac{\Lambda}{3}- \frac{k}{a^2(t,r)},
\label{Friedmann_classI}
\ee
where we have defined $\mathbb{H}$ by analogy with the Hubble parameter in $\Lambda$CDM. We point out that $\mathbb{H}$ is not the same as the metric function $H$. Continuing by analogy, the $t$- and $r$-dependent Szekeres cosmological density parameters can then be defined as
\begin{align}
	\Omega_m(t,r) &\equiv \frac{2M(r)}{a^3(t,r)\mathbb{H}^2(t,r)},\nonumber\\
	\Omega_\Lambda(t,r) &\equiv \frac{\Lambda}{3\mathbb{H}^2(t,r)},
\label{omegas_classI}
\end{align}
and
\be
	\Omega_k(t,r) \equiv \frac{-k}{a^2(t,r)\mathbb{H}^2(t,r)}\nonumber
\ee
so that Eq. (\ref{Friedmann_classI}) can be written as the sum of the three equal to $1$, as is usual in $\Lambda$CDM.

By Eq. (\ref{Szekeres_Rho_q}), we can now write Eq. (\ref{growth1_classI}) as
\be
	\ddot{\hat\delta}+2 \mathbb{H}(t,r)\,\dot{\hat\delta}-4 \pi G \rho_{q}(t,r)\,\hat\delta-\frac{2}{1+\hat\delta}\,{\dot{\hat\delta}^2}-4 \pi G \rho_q(t,r)\,\hat\delta^2=0,
\label{growth2_classI}
\ee
where we have temporarily restored the factor $8\pi G$ for clarity. Equation (\ref{growth2_classI}), in its first three terms, bears a formal similarity to the linearly perturbed FLRW equation in the density contrast. However, the Szekeres $\hat\delta$ measures deviations of $\rho$ from $\rho_q$, the quasilocal average density, as opposed to a global average as in FLRW. The $\mathbb{H}$ and $\rho_q$ terms are also dependent on $r$, and Eq. (\ref{growth2_classI}) contains two nonlinear terms that the FLRW counterpart does not. Moreover, $\hat\delta$ here is an exact quantity arising directly from an inhomogeneous metric and as such is not constrained to be smaller than $1$.

In class I, the evolution equation for $\hat\delta$ in terms of time derivatives is most useful, but we can continue the formalism in a general way that will be more useful for class II, where $a$ derivatives and density parameters are convenient. Since each surface of constant $t$ and $r$ evolves independently, we can consider fixing the value of $r$ at some $r_0$ and letting $a(t,r_0)$ serve as the time parameter for that surface. Then in terms of $a$ derivatives, Eq. (\ref{growth1_classI}) becomes
\be
	\hat\delta''+\left(\frac{\ddot{a}}{\dot{a}^2}+\frac{2}{a}\right)\hat\delta'-\frac{3M}{a^3\dot{a}^2}\hat\delta-\frac{2}{1+\hat\delta}{\hat\delta'^2}-3\frac{3M}{a^3\dot{a}^2}\hat\delta^2=0,
\label{growth_a_classI}
\ee
where $'\equiv\partial/\partial a$ at $r_0$. 

An explicit expression $a$ (parametric or otherwise) is unnecessary, since we can eliminate $\dot{a}$ and $\ddot{a}$ in favor of $\Omega_m$ and $\Omega_\Lambda$, using Eq. (\ref{Friedmann_classI}) and its time derivative. Multiplying Eq. (\ref{Friedmann_classI}) by $a^2$ and differentiating with respect to $t$ gives
\be
	\ddot{a}=-\frac{M}{a^2}+\frac{\Lambda}{3}a.
\label{addot_classI}
\ee
With this and the definitions (\ref{omegas_classI}), we obtain
\be
	\frac{\ddot{a}}{\dot{a}^2} + \frac{2}{a} = \frac{4+2\Omega_\Lambda-\Omega_m}{2a}
\ee
and
\be
	\frac{3M}{a^3\dot{a}^2}=\frac{3}{2}\frac{\Omega_m}{a^2}.
\ee
Substituting these into Eq. (\ref{growth_a_classI}), we find
\be
	\hat\delta''+\left(\frac{4+2\,\Omega_\Lambda-\Omega_m}{2a}\right)\hat\delta'-\frac{3}{2}\frac{\Omega_m}{a^2}\hat\delta-\frac{2}{1+\hat\delta}{\hat\delta'^2}-\frac{3}{2}\frac{\Omega_m}{a^2}\hat\delta^2=0.
\label{growth_a2_classI}
\ee

Finally, if we assume the cosmological parameters take on their standard meaning at every $r$, we can express them in terms of their values evaluated today (denoted by a superscript naught) and $a(t,r)$,
\be
	\Omega_m(t,r)=\frac{\Omega_m^0(r)}{\Omega_m^0(r)+\Omega_\Lambda^0(r)(a/a_0)^3+\Omega_k^0(r)(a/a_0)}
\label{OmegaM0_classI}
\ee
and
\be
	\Omega_\Lambda(t,r)=\frac{\Omega_\Lambda^0(r)(a/a_0)^3}{\Omega_m^0(r)+\Omega_\Lambda^0(r)(a/a_0)^3+\Omega_k^0(r)(a/a_0)},
\label{OmegaV0_classI}
\ee
where $\Omega_k^0(r)=1-\Omega_m^0(r)-\Omega_\Lambda^0(r)$ and $a_0=a(t_0,r)$ is the scale factor today. These relationships can then be evaluated at $r_0$ and substituted into Eq. (\ref{growth_a2_classI}).

In class I, there is an alternative method of calculating $\hat\delta$ when a model is completely specified---that is, when all the arbitrary functions of $r$ are known (see Appendix A). Instead of numerically solving the second-order differential equation in (\ref{growth1_classI}), one can compute $\hat\delta$ via the explicit functional form
\be
	\hat\delta=\left(k^2\frac{M,_r}{3M}-\frac{a,_r}{a}\right)\left(\nu,_{r}+\frac{a,_r}{a}\right)^{-1},
\label{functiondelta}
\ee
where a comma denotes partial differentiation.
%
\subsection{Growth equations in flat and curved Szekeres class II models with a cosmological constant}\label{growthEqnsClassII}
%
As discussed in Sec. \ref{GWform} above, the class II density contrast $\delta=F/H=(\rho-\bar{\rho})/\bar{\rho}$ has a similar form as in FLRW, where $\bar{\rho}(t)=6M/a^3(t)$ is the overall background average density. We also understand $\rho(t)$ as the associated FLRW model's density that underlies the exact Szekeres perturbations. This is possible since $a$ and $M$ no longer depend on $r$. Equation (\ref{growth_a2_classI}) applies equally well to class II but with $\hat\delta$ replaced by $\delta$, $a=a(t)$, and $\Omega_i=\Omega_i(t)$ for $i=\{m,\Lambda,k\}$. Now the density parameters in terms of their values today are
\be
	\Omega_m(t)=\frac{\Omega_m^0}{\Omega_m^0+\Omega_\Lambda^0\, a^3+\Omega_k^0\, a}
\ee
and
\be
	\Omega_\Lambda(t)=\frac{\Omega_\Lambda^0\, a^3}{\Omega_m^0+\Omega_\Lambda^0\, a^3+\Omega_k^0\, a},
\ee
where $\Omega_k^0=1-\Omega_m^0-\Omega_\Lambda^0$ and we have set $a_0=a(t_0)=1$. In class II we can therefore directly compare observations with the cosmological density parameters between Szekeres and $\Lambda$CDM. 

As in our previous work \cite{Ishak&Peel2012}, we find it useful to rewrite the evolution equation for $\delta$ in terms of the growth rate variable $G=(\delta/\delta_0)/a$. Noting the relations $\delta'=G+aG'$ and $\delta''=2G'+aG''$ (for the choice $\delta_0=1$), Eq. (\ref{growth_a2_classI}) for class II then becomes
\be
	G''+\left(4+\Omega_\Lambda-\frac{\Omega_m}{2}\right)\frac{G'}{a}+(2+\Omega_\Lambda-2\,\Omega_m)\frac{G}{a^2}-\frac{2}{a}\frac{(G+aG')^2}{1+aG}-\frac{3}{2}\Omega_m\frac{G^2}{a}=0,
\label{growth_G_classII}
\ee
and one can easily check that this equation reduces to the forms obtained in \cite{Ishak&Peel2012} for the special cases (i) $\Omega_\Lambda=0$ and (ii) $\Omega_m=1$, $\Omega_\Lambda=\Omega_k=0$.

%
\section{Integration of Growth History for Szekeres Models with a Cosmological Constant}\label{growthPlots}
%
The Szekeres growth equations derived in Sec. \ref{growthEqns} are integrated numerically for $\hat{\delta}$ (class I) and $G$ (class II) using a standard fourth-order Runge-Kutta routine with adaptive step size \cite{Press1992}. In class I we discuss results for a simple but inhomogeneous and nonsymmetric model representing a central void with neighboring supercluster, which is similar to Model 1 employed by \cite{Bolejko2006}. Our model is designed to match to a nearly homogeneous FLRW background model described with standard $\Lambda$CDM parameters ($\Omega_m^0=0.27$, $\Omega_\Lambda^0=0.73$, and $\mathrm{H}_0=72$ km s$^{-1}$ Mpc$^{-1}$) at a radius of $50$ Mpc and greater. We provide the algorithm for building such a model, specify the necessary arbitrary functions, and describe the resulting density structure in Appendix C. 

For class II, we compare the growth rate in Szekeres models with various combinations of cosmological parameters to the $\Lambda$CDM model with linear perturbations. The discussions for class II can be applied as well to class I if we restrict ourselves to a constant $r$-coordinate surface, where the density parameters take on the values specified for the class II model. However, as seen by Eq. (\ref{classIbeta}) of Appendix A, there are no growing modes for the spatially flat class I case, because $k=0$ forces $\beta_+$ to vanish. Since we are interested in structure growth, we therefore do not consider flat models to apply to class I.
%
\subsection{Growth history for flat Szekeres class II models}\label{growthHistFlat}
%
We find that for flat class II Szekeres and $\Lambda$CDM models with comparable values $\Omega_m^0$ and $\Omega_\Lambda^0$, the Szekeres models exhibit significantly stronger growth than their linearly perturbed $\Lambda$CDM counterparts. This difference is as expected, since the Szekeres growth equation contains exact nonlinear contributions due to inhomogeneities as well as shear and a gravitational tidal field that the standard concordance model does not.

\begin{figure}[t]
\begin{center}
\begin{tabular}{|c|c|}
\hline
\includegraphics[scale=0.49,angle=0]{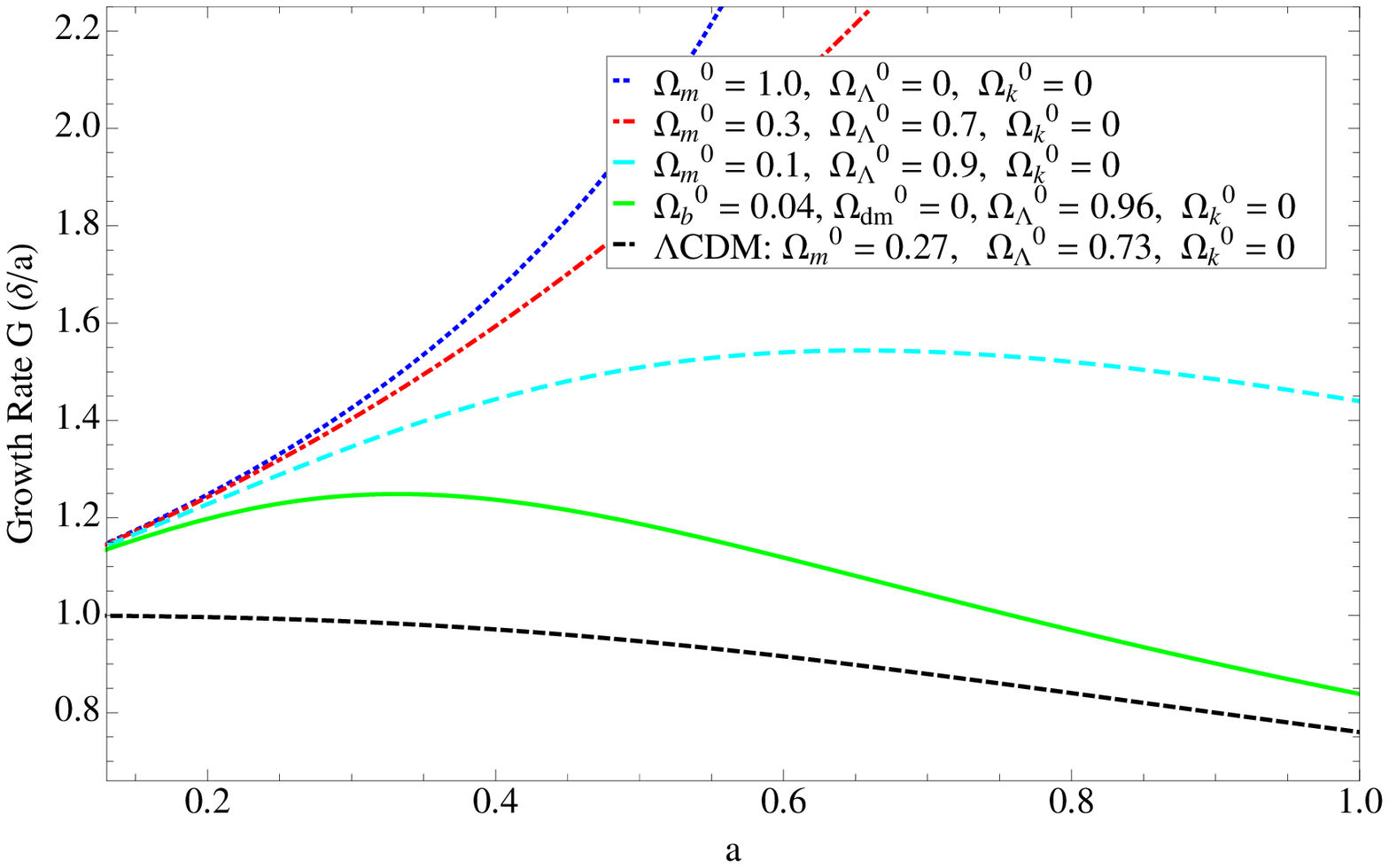}&
\includegraphics[scale=0.49,angle=0]{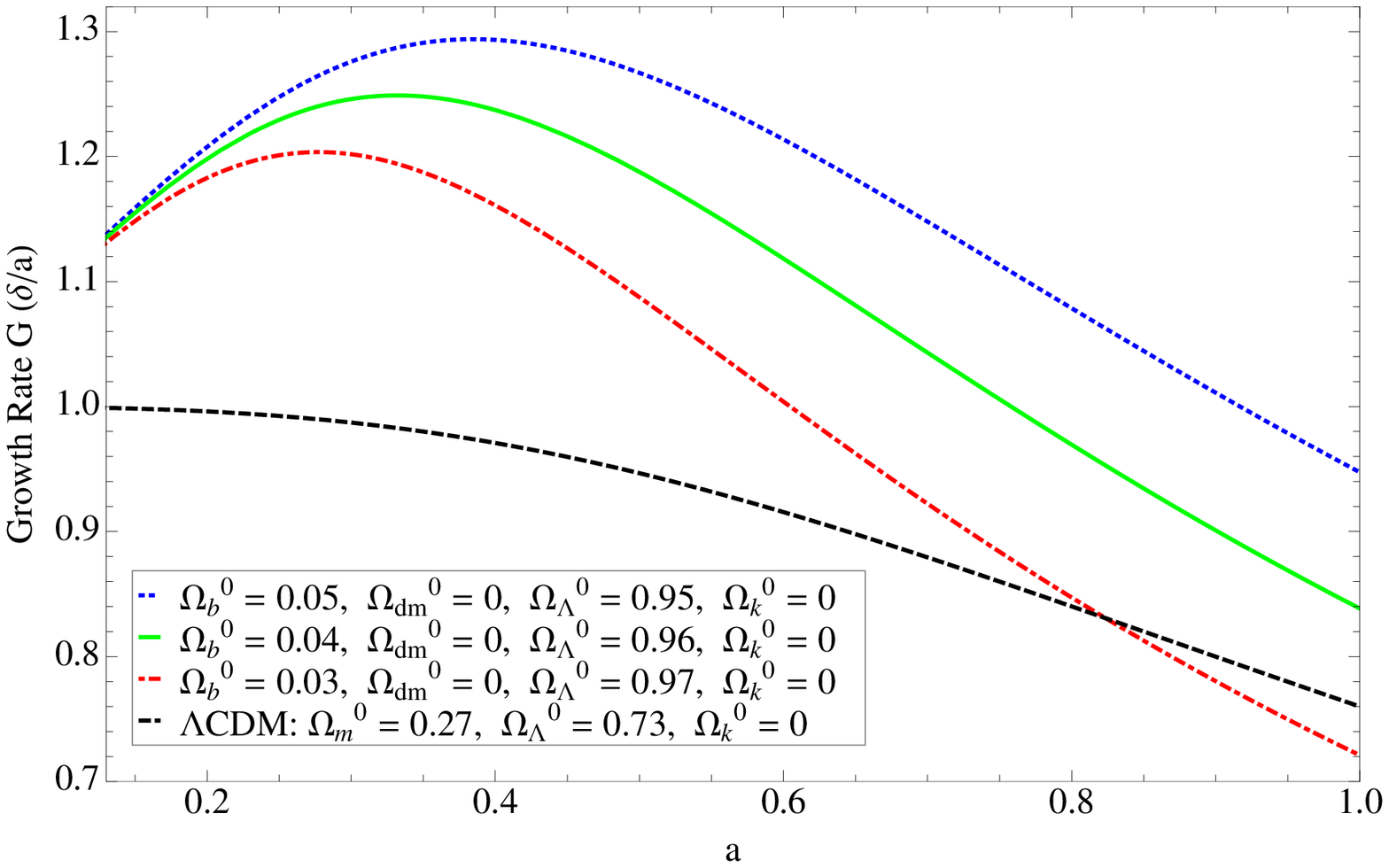}\\
\hline
\end{tabular}
\caption{\label{fig:growthFlat}
LEFT: Growth rate of large-scale structure in flat Szekeres class II models with a cosmological constant. A matter-only universe experiences the strongest growth, and curves are seen to experience significant suppression by $\Lambda$ within our cosmic history for $\Omega_m^0$ values of approximately $0.1$ and smaller. The curve with $\Omega_m^0=\Omega_b^0=0.04$ and $\Omega_\Lambda^0=0.96$ (green, solid) could provide a consistent growth history and is a good flat model candidate for comparison with $\Lambda$CDM (black, dashed), which is plotted with the standard values of $\Omega_m^0=0.27$ and $\Omega_\Lambda^0=0.73$.
RIGHT: A sample of flat class II models guided by the results of the left panel that could mimic $\Lambda$CDM (dashed) with respect to the growth of large-scale structure today. The Szekeres curves emerge at early times almost identically but begin to separate at a scale factor of about $0.15$. The $\Omega_b^0=0.05$ model is able to grow the most structure before suppression by the cosmological constant takes over. Despite differing times for the onset of suppression, the three curves experience approximately the same rate of suppression at later times, indicating the effect of a dominant cosmological constant.}
\end{center}
\end{figure}

The left panel of Fig. \ref{fig:growthFlat} shows the growth behavior over a range of $\Omega_m^0$ in flat models. For comparison, the linearly perturbed $\Lambda$CDM growth for $\Omega_m^0=\Omega_b^0+\Omega_{dm}^0=0.27$ and $\Omega_\Lambda^0=0.73$ is plotted as well. The Szekeres case with $\Omega_m^0=1.0$ and $\Omega_\Lambda^0=0$ (i.e., with an Einstein--de Sitter associated background) experiences the strongest growth with no sign of suppression within our cosmic history. For smaller values of $\Omega_m^0$, however, we begin to see suppression, which is consistent with the fact that the cosmological constant makes the expansion of the universe accelerate and gives large-scale structure less opportunity to grow. In each case, the cosmological constant contribution ultimately wins out, turning the curve over, but only for $\Omega_m \approx 0.1$ and smaller do we see the effect clearly in our cosmic history.

It is interesting that if we take into consideration the bounds $0.039\le\Omega_b^0\le 0.049$ from big bang nucleosynthesis (BBN) \cite{wmap7}, we can interpret the small matter contribution in Szekeres as being due solely to baryons (i.e., not dark matter). With the BBN limits in mind, we find that a flat Szekeres model with $\Omega_m^0=\Omega^0_b\approx0.04$ produces a growth curve that roughly mimics that of linearly perturbed $\Lambda$CDM today.  The Szekeres growth rate is larger early on than $\Lambda$CDM but still undergoes suppression to make $\delta/a$ of the two comparable at present. This is significant because considering the effect of nonlinear inhomogeneities seems to strengthen the growth of structure in a way that requires little or perhaps no dark matter. We discuss this point further in Sec. \ref{growthFactor}, where we compare the Szekeres growth to current observational data and obtain best fits for the parameters $\Omega_m^0$, $\Omega_\Lambda^0$, and $\Omega_k^0$.

In the right panel of Fig. \ref{fig:growthFlat}, we see indeed that flat Szekeres models with values of $\Omega_m^0=\Omega_b^0$ near 0.04 have growth rates comparable to $\Lambda$CDM today, with the spread arising from shifting the balance between $\Omega_m^0$ and $\Omega_\Lambda^0$. In each curve here, the growth rate emerges at early times almost identically with the others up to $a\approx0.15$, but the curves representing larger $\Omega_m^0$ values experience suppression later. As expected \cite{Peacock1999}, when $\Lambda$ dominates at late times, the growth rate becomes essentially linear with constant (negative) slope among the three cases presented.
\begin{figure}
\begin{center}
\begin{tabular}{|c|c|}
\hline
{\includegraphics[scale=0.49,angle=0]{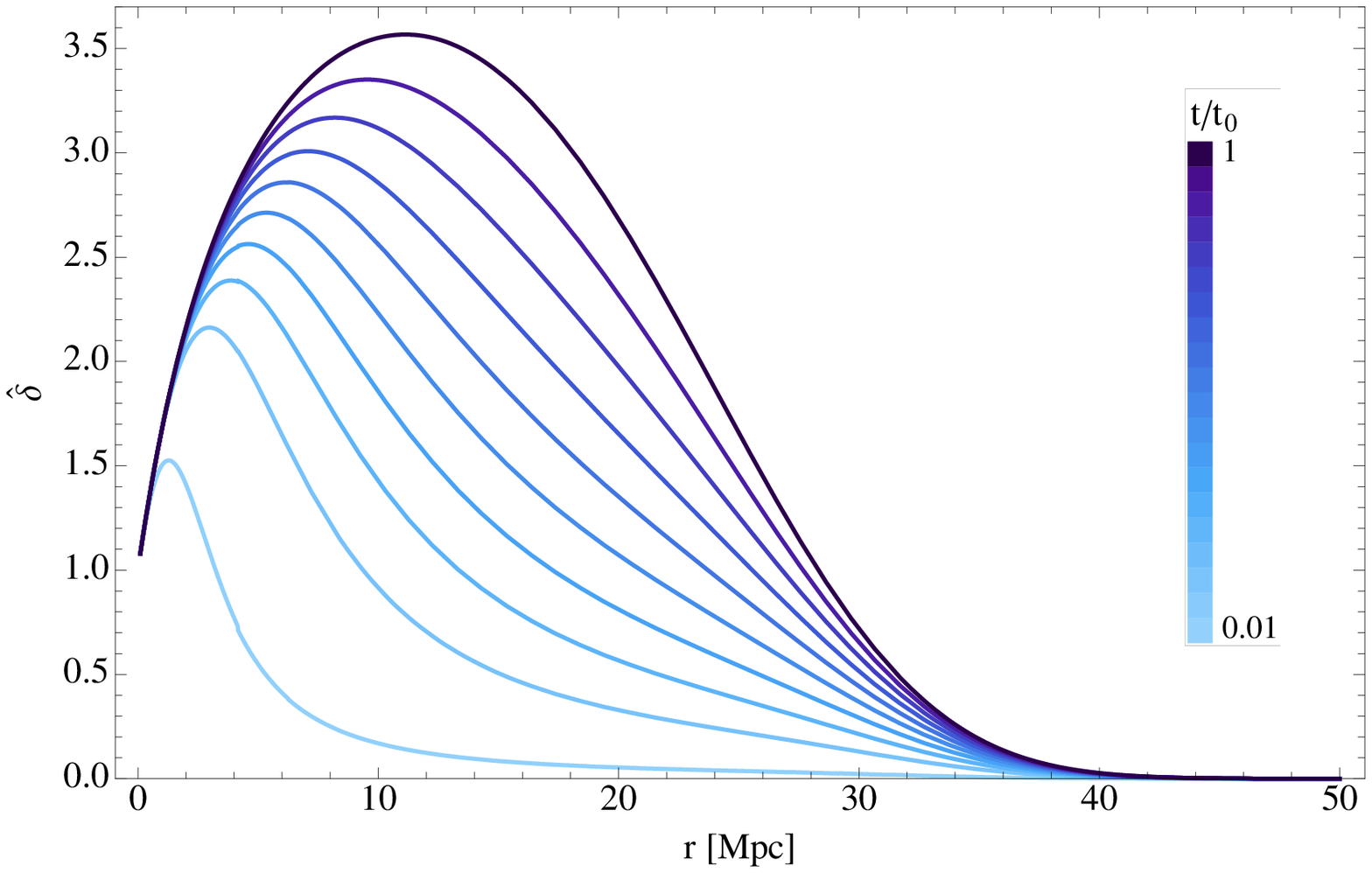}}&
{\includegraphics[scale=0.49,angle=0]{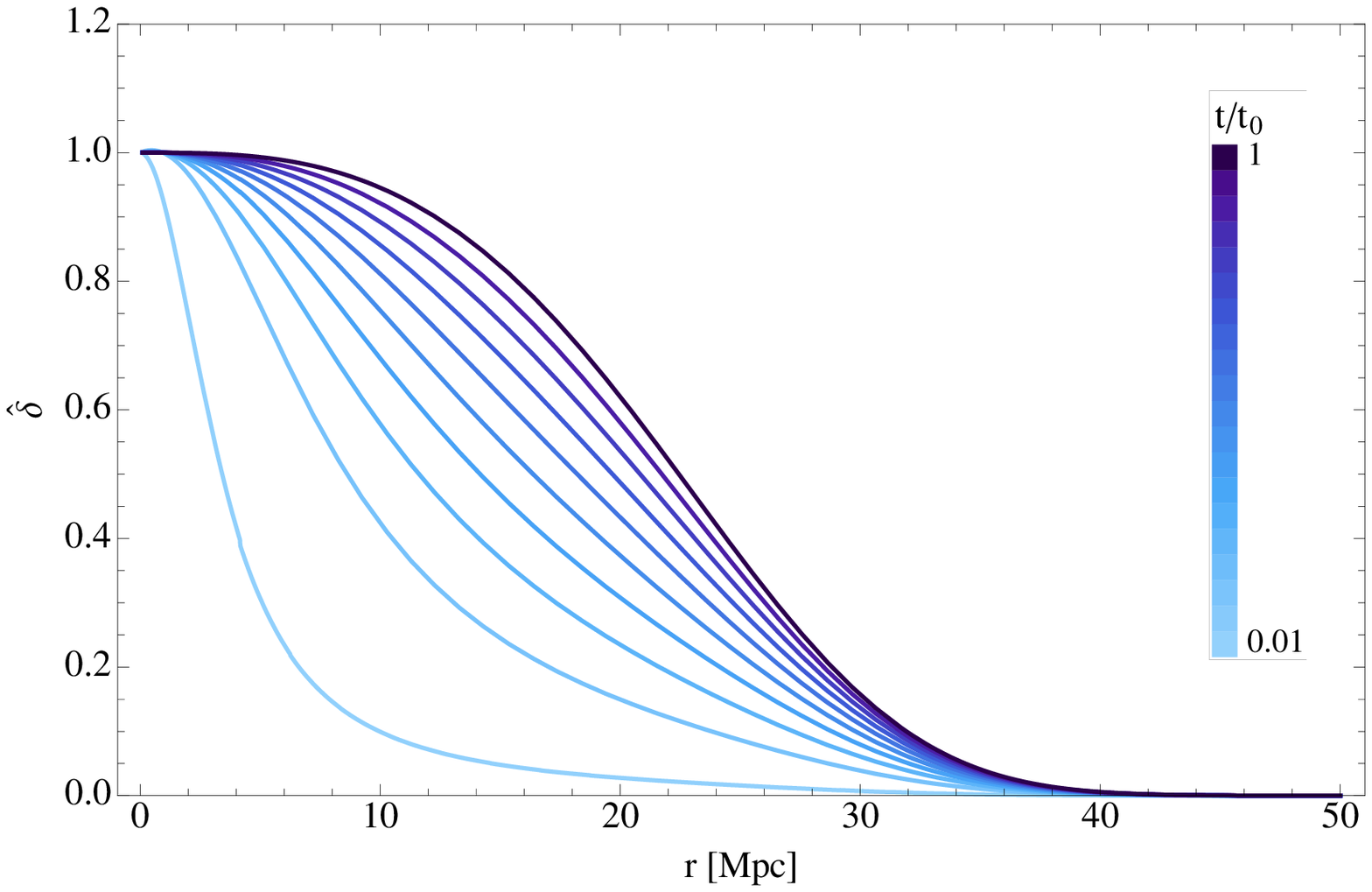}} \\
\hline  
\end{tabular}
\caption{\label{fig:classIdelta}
LEFT: Evolution of the density contrast of the class I model along a direction through the overdense region that neighbors the central void. The color gradient represents $\hat\delta$ at different times, where the present time is darkest and curves get lighter toward the past.
RIGHT: Evolution of the density contrast along a direction orthogonal to the overdensity. While in the left panel $\hat\delta$ peaks as a function of $r$, in the right panel it is a monotonically decreasing function of $r$ that never exceeds $1$. In both panels, $\hat\delta$ vanishes beyond $50$ Mpc, which we expect due to our model matching to an almost FLRW background at that radius. The curves are evenly spaced in time, and the density contrast in both directions is seen to decrease toward the past, indicating a smoothing out of the region.
}
\end{center}
\end{figure}
%
\subsection{Growth history for curved Szekeres models}\label{growthHistCurved}
%
While the global spatial curvature of the universe has been well constrained to be negligibly small using the $\Lambda$CDM model \cite{wmap7} this is not the case when using inhomogeneous cosmological models \cite{bolcmbdist}. As we will see in Sec. \ref{growthFactor}, the zero curvature result using FLRW may be biased due to assumptions that do not hold in the more general Szekeres models. We compare there class II models to growth data and find a significant curvature component, which is absent in the $\Lambda$CDM model. 

\subsubsection{Class I}
The evolution of the density contrast for our class I model (described in Appendix C) is shown in Fig. \ref{fig:classIdelta}. Though the model matches (at some given $r$) to a nearly homogenous $\Lambda$CDM background with zero curvature ($\Omega_m^0=0.27$, $\Omega_{\Lambda}^0=0.73$), the Szekeres region itself has a nonzero curvature ($k=-1$). The left panel shows $\hat\delta$ as a function of $r$ in a direction passing through the overdensity adjacent to the central void. There is a prominent maximum today at $r\approx11$ Mpc, and this peak diminishes toward the past as well as moves toward $r=0$. The density contrast profile in this direction achieves the largest values compared with other directions, which is consistent with the expectation that the density deviates most from the background through regions of larger structure. However, we must keep in mind that the background density at a given $r$ is a quasi-local average density $\rho_q$ defined as the average within the volume bounded by the 2-surface of constant $r$ and $t$. The maximum of $\hat\delta$ therefore need not coincide with the maximum of $\rho$ for some value of $t$ (see Fig. \ref{fig:classIrho}). The right panel plots $\hat\delta$ along a direction orthogonal to the overdensity. The density contrast here is a monotonically decreasing function of $r$ for all times shown, and the density deviation from the background is seen to lessen toward the past. It is clear that less structure lies along this direction compared to that of the left panel. That our model matches to an almost FLRW background (as described at the beginning of Sec. \ref{growthPlots}) at $50$ Mpc is apparent in both plots, since $\hat\delta$ approaches zero there. We find here that the growth in a Szekeres class I model is up to several times stronger than that of the linearly perturbed $\Lambda$CDM model ($\delta \ll 1$), a result consistent with previous works \cite{Bolejko2007,Ishak&Peel2012}.  

\subsubsection{Class II}
\begin{figure}
\begin{center}
\begin{tabular}{|c|c|}
\hline
{\includegraphics[scale=0.49,angle=0]{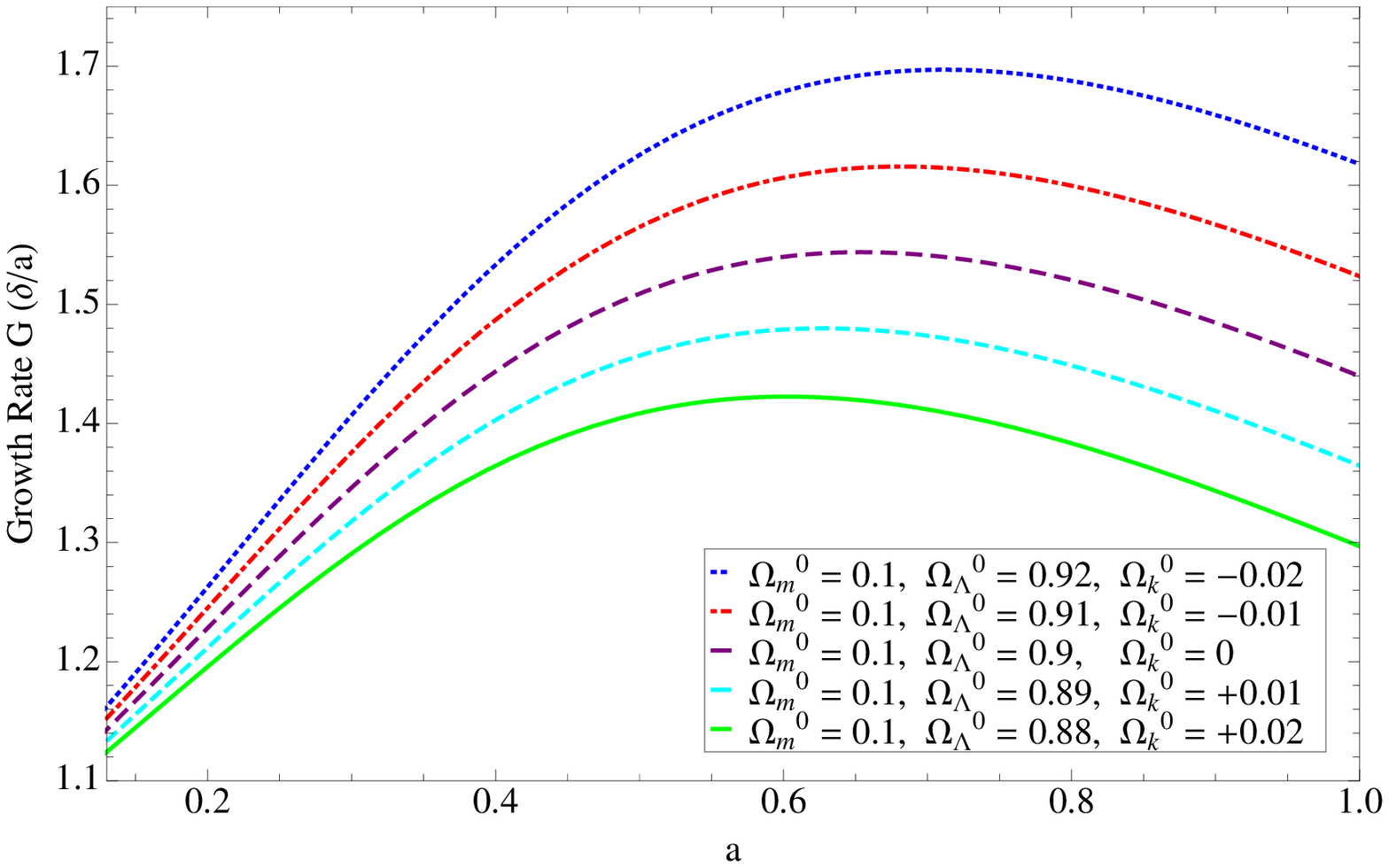}}&
{\includegraphics[scale=0.49,angle=0]{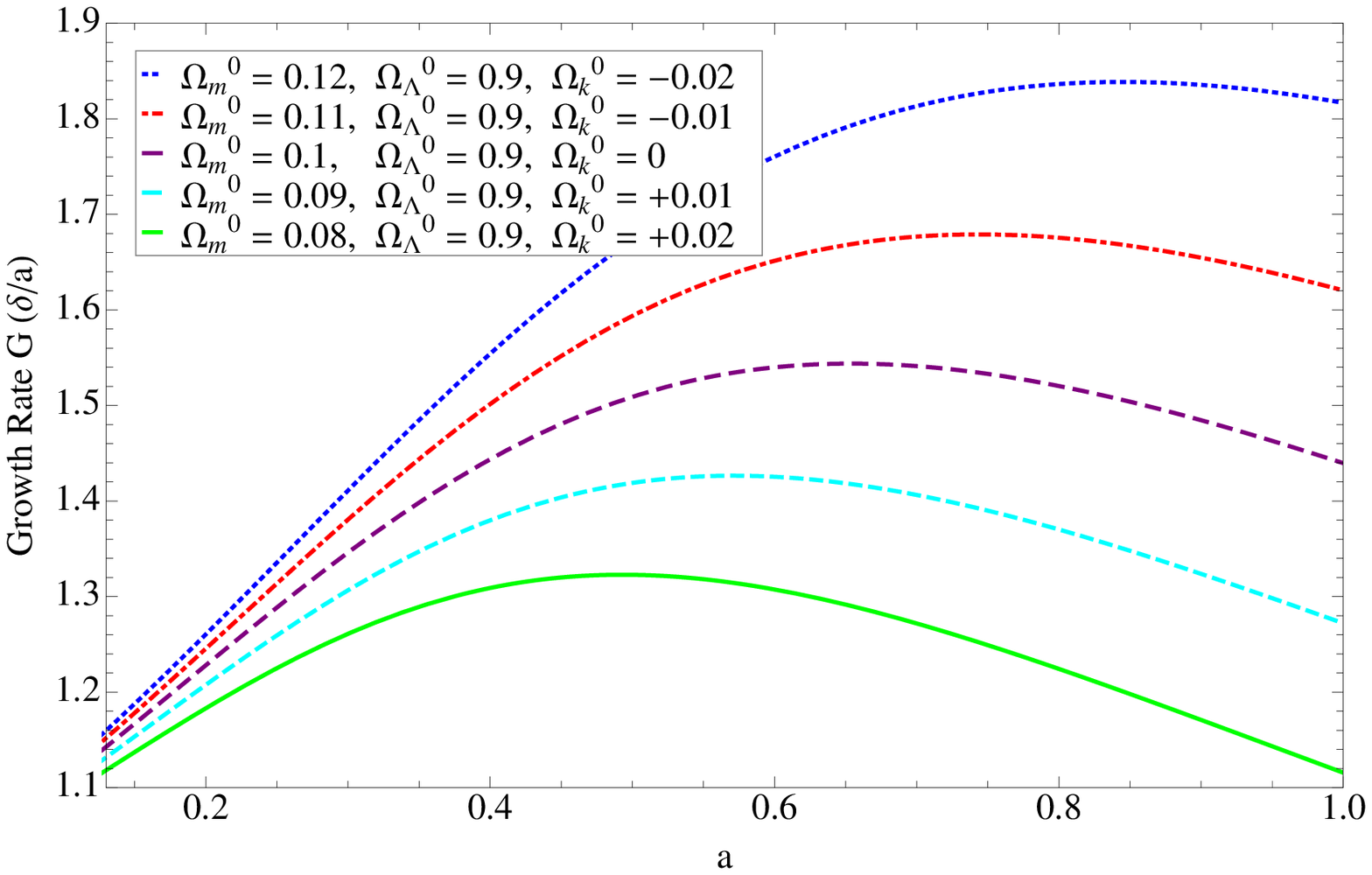}} \\
\hline  
\end{tabular}
\caption{\label{fig:growthCurv}
LEFT: Class II Szekeres growth histories for $\Omega_m^0=0.1$ and varying $\Omega_\Lambda^0$ and $\Omega_k^0$. Closed models ($\Omega_k^0 < 0$) experience stronger growth than open models ($\Omega_k^0 > 0$) overall. The flat case lies in between the two, and all models have stronger growth compared to standard $\Lambda$CDM (not shown; see Fig. \ref{fig:growthFlat}). All the Szekeres curves undergo the same characteristic suppression by $\Lambda$ after attaining their maximum growth rates (which happens later the smaller $\Omega_k^0$ is) so that the slopes becomes approximately the same across the different curves.
RIGHT: Szekeres growth histories for $\Omega_\Lambda^0=0.9$ and varying $\Omega_m^0$ and $\Omega_k^0$. Closed models, which have larger matter density, again experience stronger growth than open models with the flat case lying in between. The spread in the curves is slightly more pronounced here than for fixed $\Omega_m^0$ over the same range of $\Omega_k^0$ values, indicating the effect that small changes in the matter content can have. All models have stronger growth at all times than $\Lambda$CDM (not shown; see Fig. \ref{fig:growthFlat}) and ultimately get suppressed by $\Lambda$ before $a$ reaches 1.
}
\end{center}
\end{figure}

The inclusion of curvature into the growth for class II provides another degree of freedom that allows for a variety of different growth histories. To see clearly the effect of all three components (matter, curvature, and $\Lambda$) on $\delta/a$, we fix one component at a time and vary the other two. For example, in Fig. \ref{fig:growthCurv} we first set $\Omega_m^0=0.1$ and vary $\Omega_\Lambda^0$ and $\Omega_k^0$. For the same matter content, closed models ($\Omega_k^0<0$) experience stronger overall growth than open ($\Omega_k^0>0$) models, with the flat case falling in between. At early times, curvature seems to be the determining factor in the growth rate, since it is the curve with $\Omega_k^0=-0.02$ (and also the largest value of $\Omega_\Lambda^0$) that rises quickest early on. But as expected, later the cosmological constant effects its characteristic suppression so that the evolution of each curve is approximately the same after it reaches its maximum.

Next, the effects of fixing $\Omega_\Lambda^0$ at 0.9 and varying $\Omega_m^0$ and $\Omega_k^0$ are shown in the right panel of Fig. \ref{fig:growthCurv}. This plot spans the same curvature range as the one to the left, yet here $G$ exhibits more pronounced variation. In this case, closed models correspond to having more matter, while open models to having less. The trend in the curves therefore matches the expectation that more matter in a closed universe will have a stronger growth rate of structure than in the opposite case. Again, $\Lambda$ suppresses the growth uniformly after the curves turn over, but now the maxima occur at very different times. We also note that the presence of nonzero $\Omega_k^0$ causes the curves to diverge from each other quite early---even earlier than $a\approx0.1$. Flat models for the parameters in Fig. \ref{fig:growthFlat}, by contrast, evolve almost identically up to scale factors of about $0.15$.

As we will see in Sec. \ref{growthFactor}, we can find parameters such that the Szekeres growth history exactly matches that of $\Lambda$CDM. The parameters that accomplish this are $\Omega_m^0=0.11$, $\Omega_\Lambda^0=0.71$, and $\Omega_k^0=0.18$. In Fig. \ref{fig:growthFlat}, its $G$ curve would be indistinguishable from $\Lambda$CDM (black, dashed). In this case, if we follow the predictions of BBN, the need for dark matter cannot be fully eliminated, but the amount needed to achieve the same growth as $\Lambda$CDM is reduced by a factor of about $3$.
 Also, here the dark energy density (or cosmological constant) contribution is nearly the same as for $\Lambda$CDM. The Szekeres $\Omega_k^0$ is significantly higher than that of $\Lambda$CDM (which is consistent with zero). The $\Omega$'s in the Szekeres models do represent fractions of the total energy content of the universe for different constituents, as defined in the Friedmann equations, which are formally identical in both the Szekeres and $\Lambda$CDM models. However, the energy densities do not have strictly the same connections to physical quantities in both cases. For example, the Hubble parameter in FLRW is proportional to the expansion scalar $\Theta$, but in Szekeres there is also a dependence on the metric function $H$ (see Sec. \ref{scalars}).

Based on the results of integrating the growth equations, we find that class II Szekeres models do have the ability to mimic the growth rate of large-scale structure in $\Lambda$CDM today for appropriately chosen values of the cosmological parameters. In particular, the parameters $\Omega_m^0=\Omega_b^0\approx0.04$ and $\Omega_\Lambda^0\approx 0.95$ (so that the universe has a small negative curvature of order $10^{-2}$) do a fair job qualitatively, and interpreting the matter content as being solely due to baryons supports the idea that Szekeres models can provide a consistent growth history with little or no dark matter. Alternatively, the parameters $\Omega_m^0=0.11$, $\Omega_\Lambda^0=0.71$, and $\Omega_k^0=0.18$ exactly reproduce the $\Lambda$CDM growth history, but seem to require some dark matter and large $\Omega_k^0$. We present a quantitative argument for smaller $\Omega_m^0$ in Szekeres than $\Lambda$CDM based on fits to the data for the growth factor in Sec. \ref{growthFactor}.

Finally, we also observe that the same trends in the growth rate hold in general for FLRW models as they do for their Szekeres counterparts. In other words, closed universes with larger matter contributions experience stronger growth overall than open universes with less matter. Flat universes lie in between. The effect of a cosmological constant in $\Lambda$CDM is also to suppress the growth rate equally at late times when it becomes dominant, regardless of the growth rate at earlier times, a well-known and expected result, e.g.  \cite{Peacock1999}.
\begin{figure}
\begin{center}
\begin{tabular}{|c|c|}
\hline
{\includegraphics[scale=0.49,angle=0]{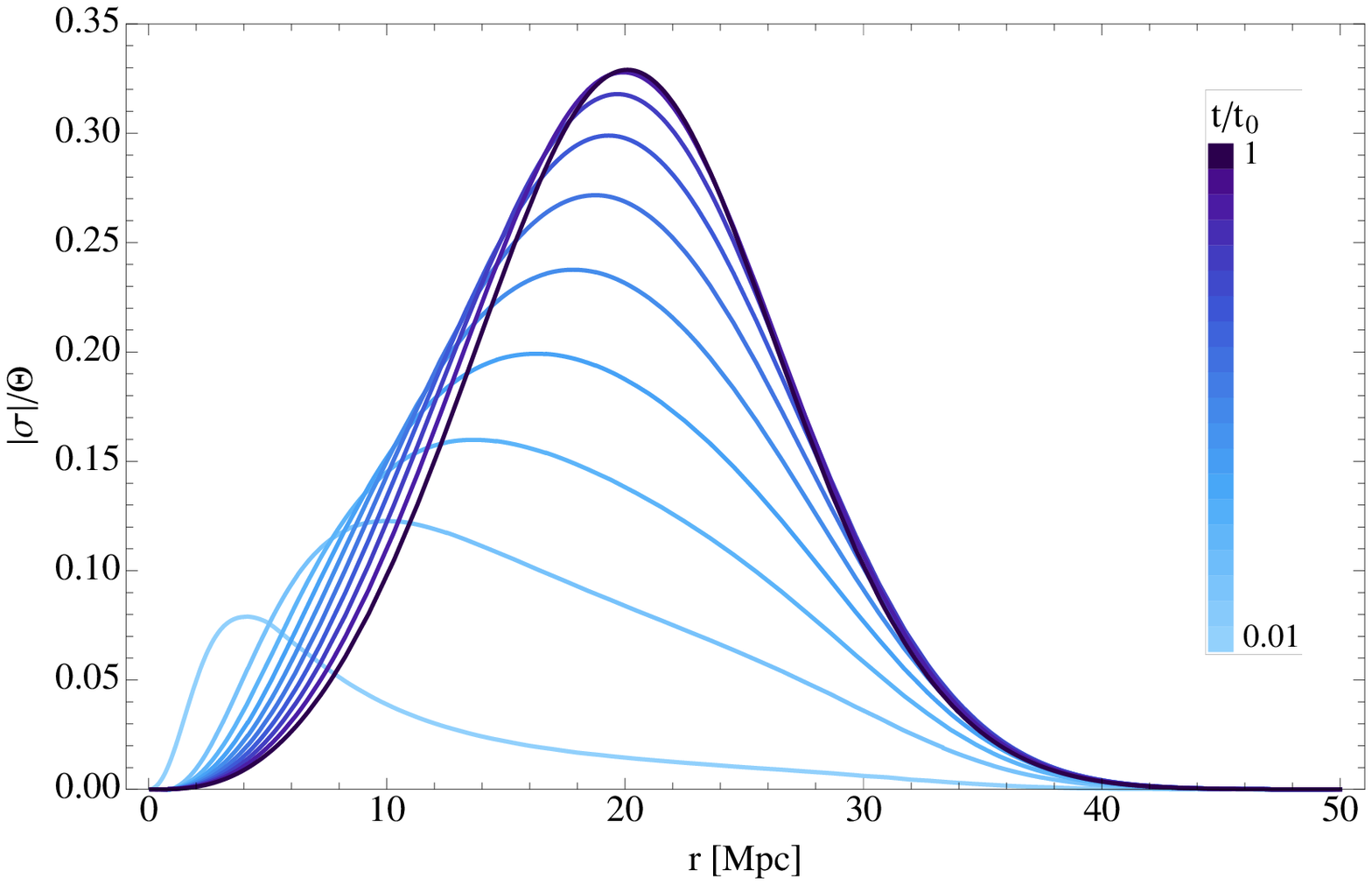}}&
{\includegraphics[scale=0.49,angle=0]{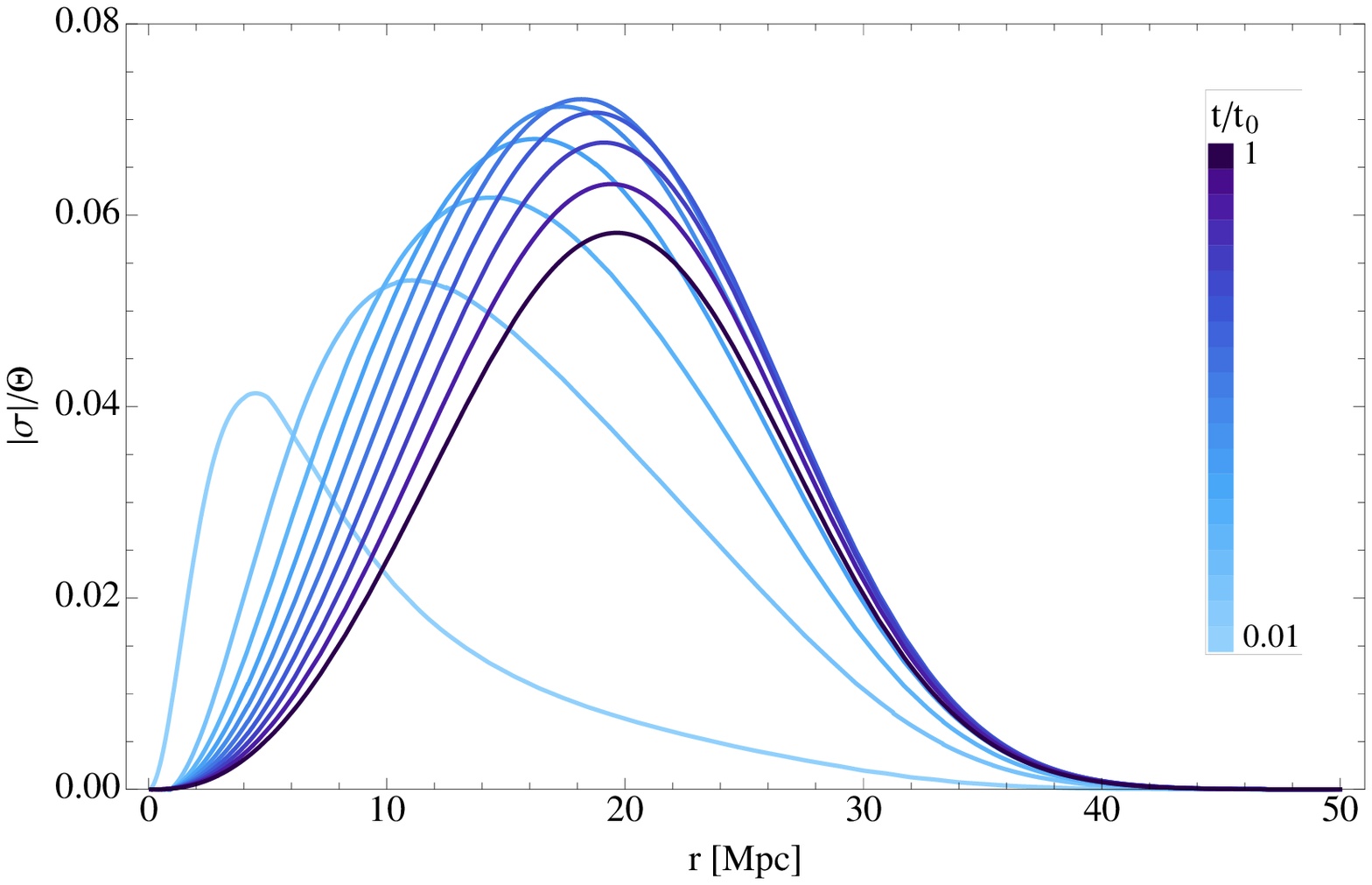}} \\
\hline  
\end{tabular}
\caption{\label{fig:classIsig}
LEFT: Evolution of $|\sigma/|\Theta$ in the class I model along a direction through the overdense region that neighbors the central void. The color gradient represents $|\sigma/|\Theta$ at different times where the present time is darkest and curves get lighter toward the past. The maximum is achieved today at $r\approx20$ Mpc, and the shear-to-expansion ratio diminishes continuously toward the past with the peak moving toward $r=0$.
RIGHT: Evolution of $|\sigma/|\Theta$ in the class I model along a direction orthogonal to the overdensity. The curves in this case first increase as we move backward in time before ultimately decreasing as in the left plot. The curves are evenly spaced in time, and so $|\sigma/|\Theta$ is changing at a slower rate today than it was in the past. We expect the shear to tend toward zero in the past as the model smooths out, and looking through the structure in both directions reveals that this is the case.
}
\end{center}
\end{figure}
%
%
\section{Cosmological Evolution of Shear and Tidal Field Scalars in Szekeres Models Including Curvature and Cosmological Constant}\label{scalars}
%
We now consider the time evolution of physical scalar quantities in the Szekeres models that influence the gravitational clustering and structure growth. Our discussion here is brief, since much of the work was carried out in our earlier paper \cite{Ishak&Peel2012}, and the inclusion of $\Lambda$ in the expressions for the shear scalar and other quantities is straightforward. In class I, due to the $r$ dependence of $a$, it is natural to express these quantities in terms of $\hat\delta$, while in class II we convert to $G$ as in our previous work.
%
\subsection{Shear and expansion}
%
The first quantity of interest is the squared shear scalar defined by
\be
	\sigma^2 \equiv \sigma^\mu{}_\nu\sigma^\nu{}_\mu,
\ee
where the mixed tensor components are given by \cite{typo1}
\be
	2\sigma^{\tilde{x}}{}_{\tilde{x}}=2\sigma^{\tilde{y}}{}_{\tilde{y}}=-\sigma^r{}_r=-\frac{2}{3}\frac{\dot{H}}{H}.
\ee
We can write then for class I,
\be
	\sigma^2=\frac{2}{3}\frac{\dot{\hat\delta}^2}{(1+\hat\delta)^2},
\ee
and for class II, after some algebra,
\be
	\sigma^2=\frac{2}{3}\,\mathbb{H}_0^2\left(\frac{\Omega_m^0}{a}+\Omega_\Lambda^0 a^2+\Omega_k^0\right)\left(\frac{G+aG'}{1+aG}\right)^2,
\label{shear}
\ee
where $\mathbb{H}_0$ is the Hubble parameter today and $a_0=1$. As we noted in previous work \cite{Ishak&Peel2012}, nonzero shear augments the gravitational attraction, resulting in stronger collapse than would be produced by the energy density alone.

We next give the expression for the expansion scalar $\Theta$ in order to examine early- and late-time divergences of $|\sigma|/\Theta$. The expansion scalar is defined as
\be
	\Theta \equiv u^\mu{}_{;\mu} = 3\,\frac{\dot{a}}{a}+\frac{\dot{H}}{H},
\ee
which we find for class I is simply
\be
	\Theta=3\,\frac{\dot{a}}{a}-\frac{\dot{\hat\delta}}{1+\hat\delta},
\ee
and for class II can be written
\be
	\Theta = \mathbb{H}_0\left(\frac{3}{a}+\frac{G+aG'}{1+aG}\right)\sqrt{\frac{\Omega_m^0}{a}+\Omega_\Lambda^0 a^2+\Omega_k^0}.
\label{Theta}
\ee
\begin{figure}
\begin{center}
\begin{tabular}{|c|c|}
\hline
{\includegraphics[scale=0.49,angle=-0]{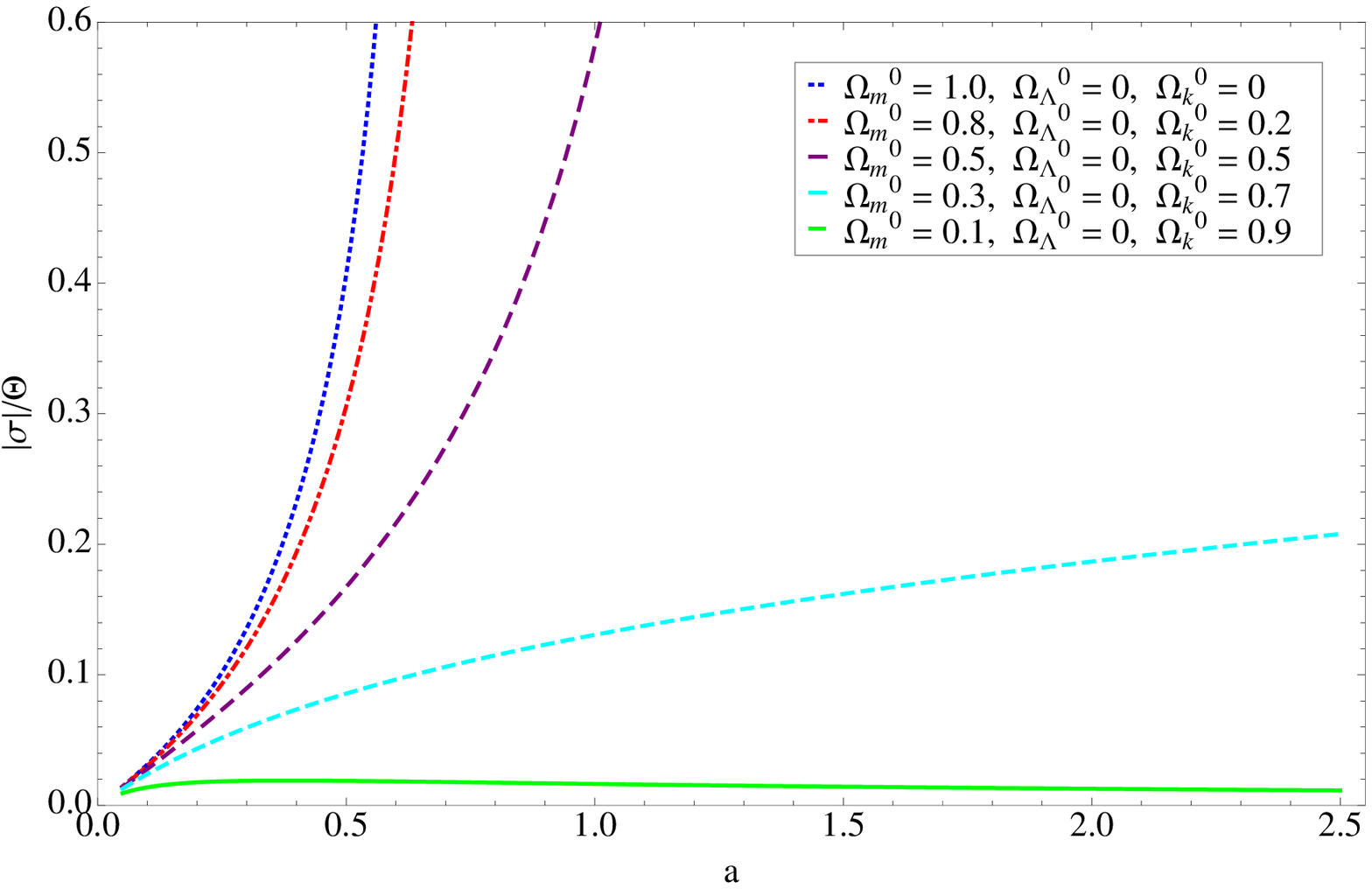}}&
{\includegraphics[scale=0.495,angle=-0]{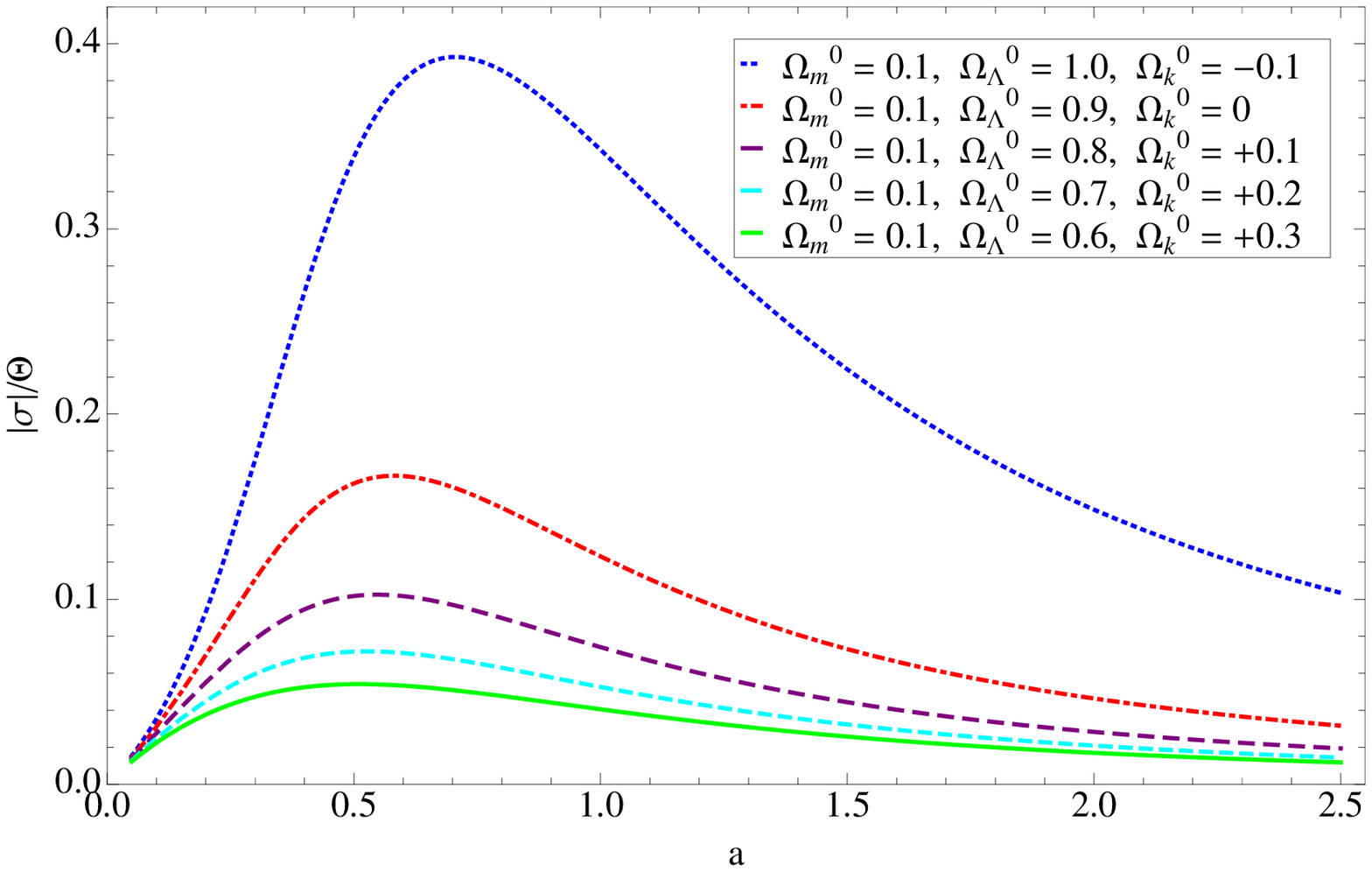}} \\
\hline  
\end{tabular}
\caption{\label{fig:SigOverThPlot}
LEFT: Shear scalar over the expansion for class II Szekeres models with no cosmological constant. We see $|\sigma|/\Theta$ going to zero at early times and potentially diverging as $a$ becomes large depending on the amount of matter and curvature. The flat matter-only case diverges most strongly, and the curves grow without bound down to $\Omega_m^0\approx 0.4$, below which we see the curves turn over and remain finite in the future. RIGHT: Shear scalar over the expansion for class II models with a cosmological constant and fixed $\Omega_m^0=0.1$. $|\sigma|/\Theta$ in all these cases goes to zero for small $a$ as in the models of the left panel. The curves reach a maximum between $0.4<a<0.8$ and then tend to zero again as $a$ becomes large with no divergences. In models where $\Omega_\Lambda^0>0.9$ (positively curved), the ratio $|\sigma|/\Theta$ attains a higher maximum at later $a$ than those where $\Omega_\Lambda^0<0.9$ (negatively curved), and the flat case falls in between.}
\end{center}
\end{figure}

We plot $|\sigma|/\Theta$ for the class I model described in Appendix C along a direction passing through the overdensity (left) and one orthogonal to it (right) in Fig. \ref{fig:classIsig}. The left plot has its maximum curve today with a peak at $r\approx20$ Mpc, and the curves there decrease with peaks shifting toward smaller $r$ as we move toward the past. In the right plot, where we are looking through less structure, the peak in $|\sigma|/\Theta$ today also occurs at about $20$ Mpc, but toward the past now the curves first rise before ultimately decreasing. In both cases, the ratio of shear to expansion vanishes at $r\ge50$ Mpc, which is expected due to the matching: all FLRW models have zero shear, and the one we match to has nonzero (positive) expansion. Also, since the curves are evenly spaced in time, we see that $|\sigma|/\Theta$ was changing faster in the past than it is today for both directions, and nowhere does it diverge for the times and ranges plotted. 

The left panel of Fig. \ref{fig:SigOverThPlot} shows $|\sigma|/\Theta$ for various class II models without a cosmological constant. These and the following class II figures have been extended beyond $a=1$ to see the trends in the curves more clearly. The ratio of shear to expansion goes to zero as $a$ approaches zero, indicating an isotropic initial singularity, and it potentially grows without bound in the future depending on the values of the matter and curvature density parameters. We can also see this by taking limits of the expression for $|\sigma|/\Theta$. The matter-only case grows most rapidly and shows no signs of turning over down to $\Omega_m^0$ values of approximately $0.4$. Below this value, where the models are strongly negatively curved, the shear and expansion evolve in proportion to each other, and the ratio does not diverge. The right panel shows $|\sigma|/\Theta$ for class II models with a cosmological constant and $\Omega_m^0$ fixed at $0.1$. Again, all curves shown go to zero for small $a$ but now also tend to zero as $a$ becomes large, attaining a maximum between scale factor values $0.4<a<0.8$. Curves for models with $\Omega_\Lambda^0>0.9$ (positively curved) reach higher maxima and at a later $a$ value than those where $\Omega_\Lambda^0<0.9$ (negatively curved), and the flat case falls in between.

We note that the explicit $\Omega_\Lambda^0$ dependence cancels out in the expression for $|\sigma|/\Theta$ in class II, which makes it identical to that previously found for Szekeres models without a cosmological constant. However, $\Omega_\Lambda^0$ is bound up in the equation for $G$, so nonzero $\Omega_\Lambda^0$ still affects the growth history and therefore $|\sigma|/\Theta$ as well. This same observation applies to the density $\rho$ and the tidal gravitational field scalar $E$ that we examine next. 
%
\subsection{Density and tidal gravitational field}
%
As seen in Sec. \ref{growthEqns}, the equation for the density in class I is given by
\be
	\rho=\frac{6M}{a^3}(1+\hat\delta),
\ee
which can be written in terms of $G$ for class II as
\be
	\rho=\frac{3\Omega_m^0}{a^3}\mathbb{H}_0^2(1+aG),
\label{rhoG}
\ee
where we have used the definition of $\Omega_m$. 

\begin{figure}
\begin{center}
\begin{tabular}{|c|c|}
\hline
{\includegraphics[scale=0.49,angle=0]{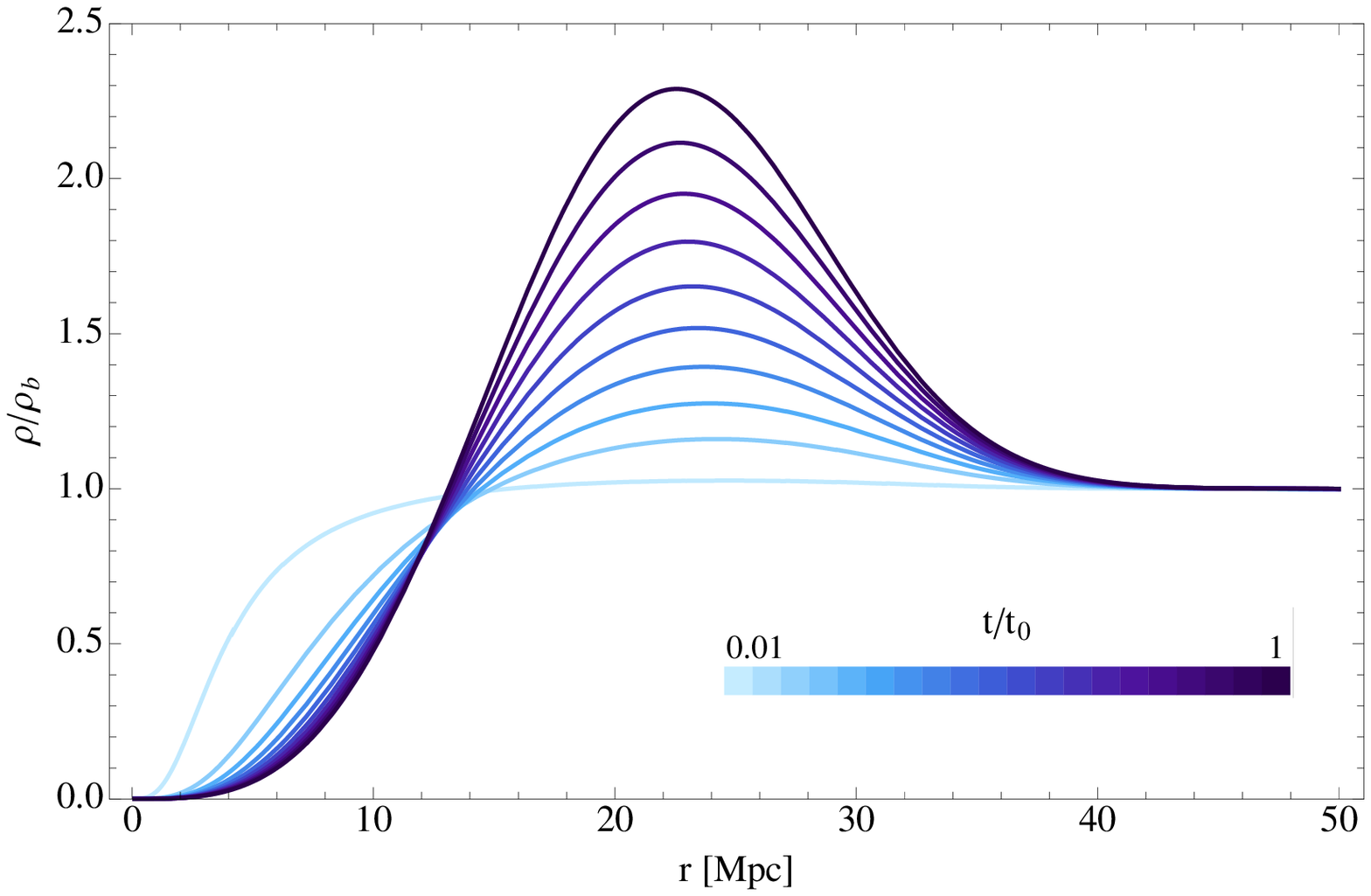}}&
{\includegraphics[scale=0.49,angle=0]{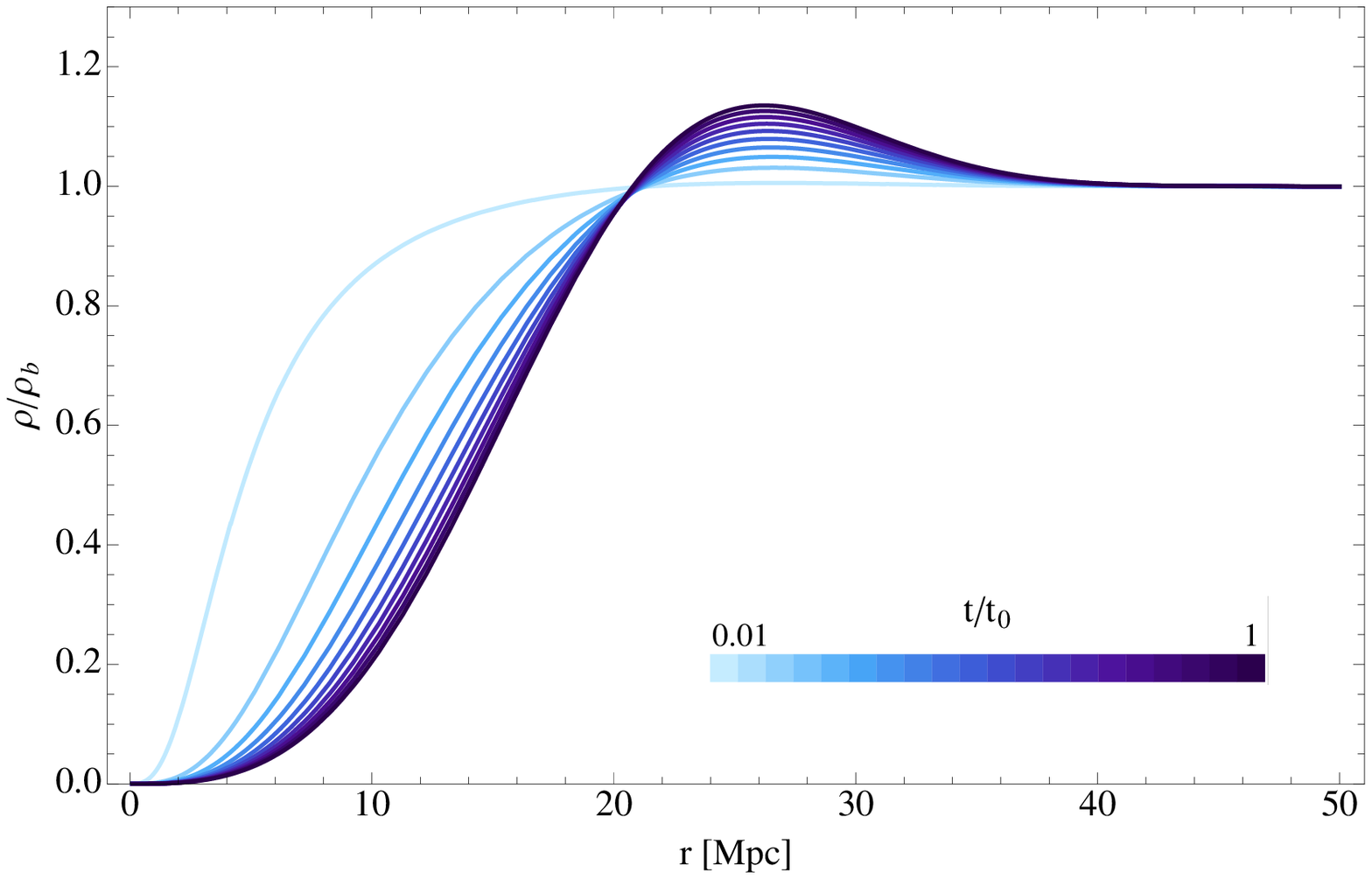}} \\
\hline  
\end{tabular}
\caption{\label{fig:classIrho}
LEFT: Evolution of the density (normalized by the almost FLRW background density $\rho_b$) of the class I model along a direction through the overdense region that neighbors the central void. The color gradient represents $\rho/\rho_b$ at different times where the present time is darkest and curves get lighter toward the past. This is the region with maximum density in the model, but it smooths out (i.e., becomes more homogeneous) significantly in the past, as expected.
RIGHT: Evolution of the normalized density along a direction orthogonal to the supercluster. The maximum overdensity attained here is about half that of the left panel, while the underdense regions in both have approximately the same profiles at corresponding times. The model also smooths out in this direction toward the past. In both panels, the curves are evenly spaced in time, from which we can infer that though the rate of growth of the overdense region is undiminished up to the present, the rate of growth in the regions with less structure seems to decrease with time. The central void's volume continues to grow in both directions over time in order to compensate the growth of the overdensity.
}
\end{center}
\end{figure}

In Fig. \ref{fig:classIrho} we plot the density, normalized by the almost FLRW background density $\rho_b$, of our void plus supercluster class I model as a function of $r$, again looking through (left panel) and orthogonal to (right panel) the overdensity. As expected, the region with the supercluster has the highest density today, and in both panels the deviation from the background becomes smaller toward the past. If viewed in terms of an associated density contrast, where we consider density differences with respect to the limiting FLRW value, then we see the quantity $\rho/\rho_b-1$ go to zero as we move backward in time. That is, the overdensity and the void both get smaller and smooth out, leaving just a void with ever-shrinking volume. It is interesting that the rate of growth of the overdense region is undiminished up to the present, while the rate of growth of the regions with less structure seem to decrease with increasing time. However, the central void's volume continues to grow in both directions over time in order to compensate for the growth of the overdensity.
\begin{figure}
\begin{center}
\begin{tabular}{|c|c|}
\hline
{\includegraphics[scale=0.5,angle=-0]{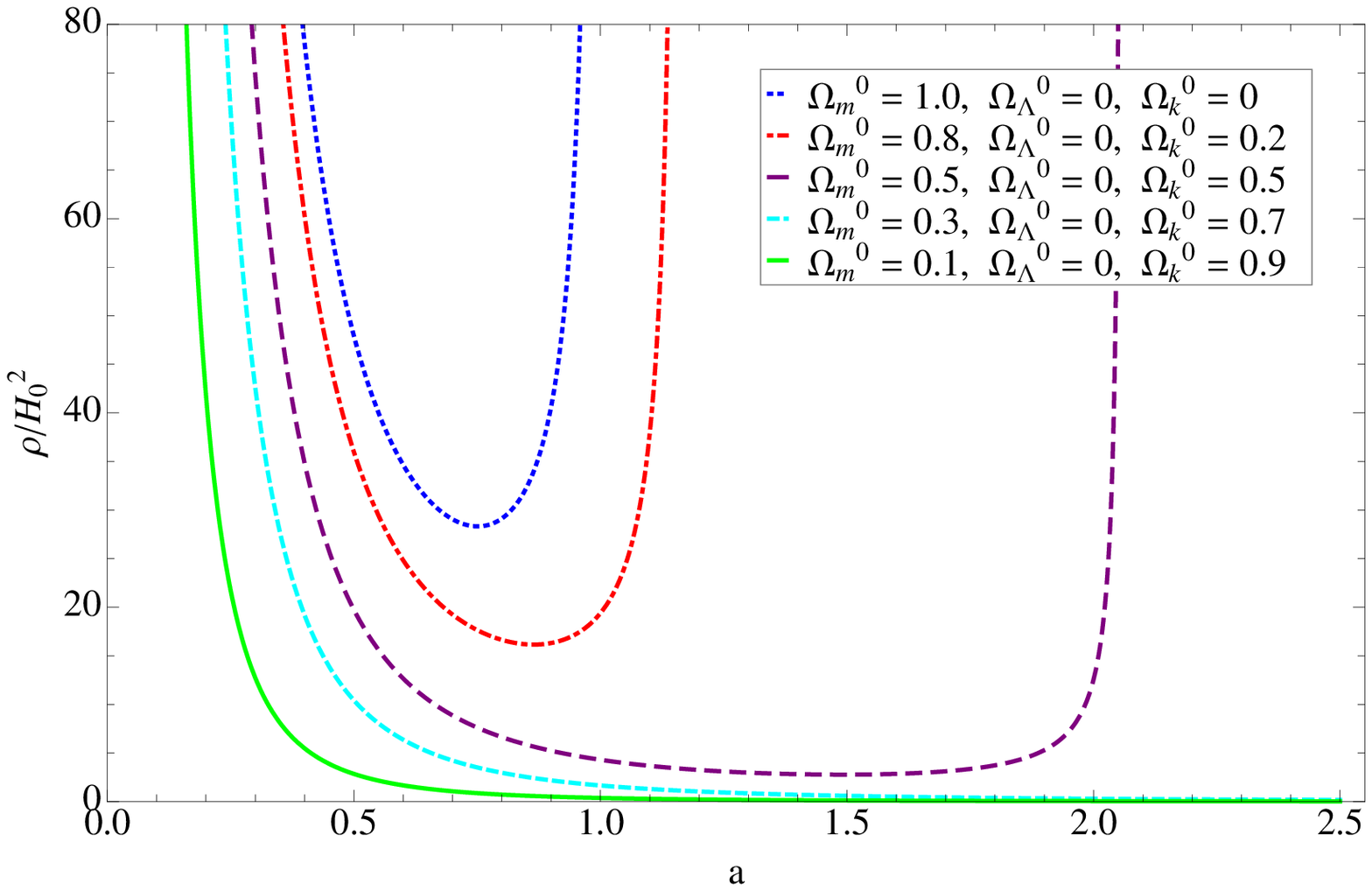}}&
{\includegraphics[scale=0.49,angle=-0]{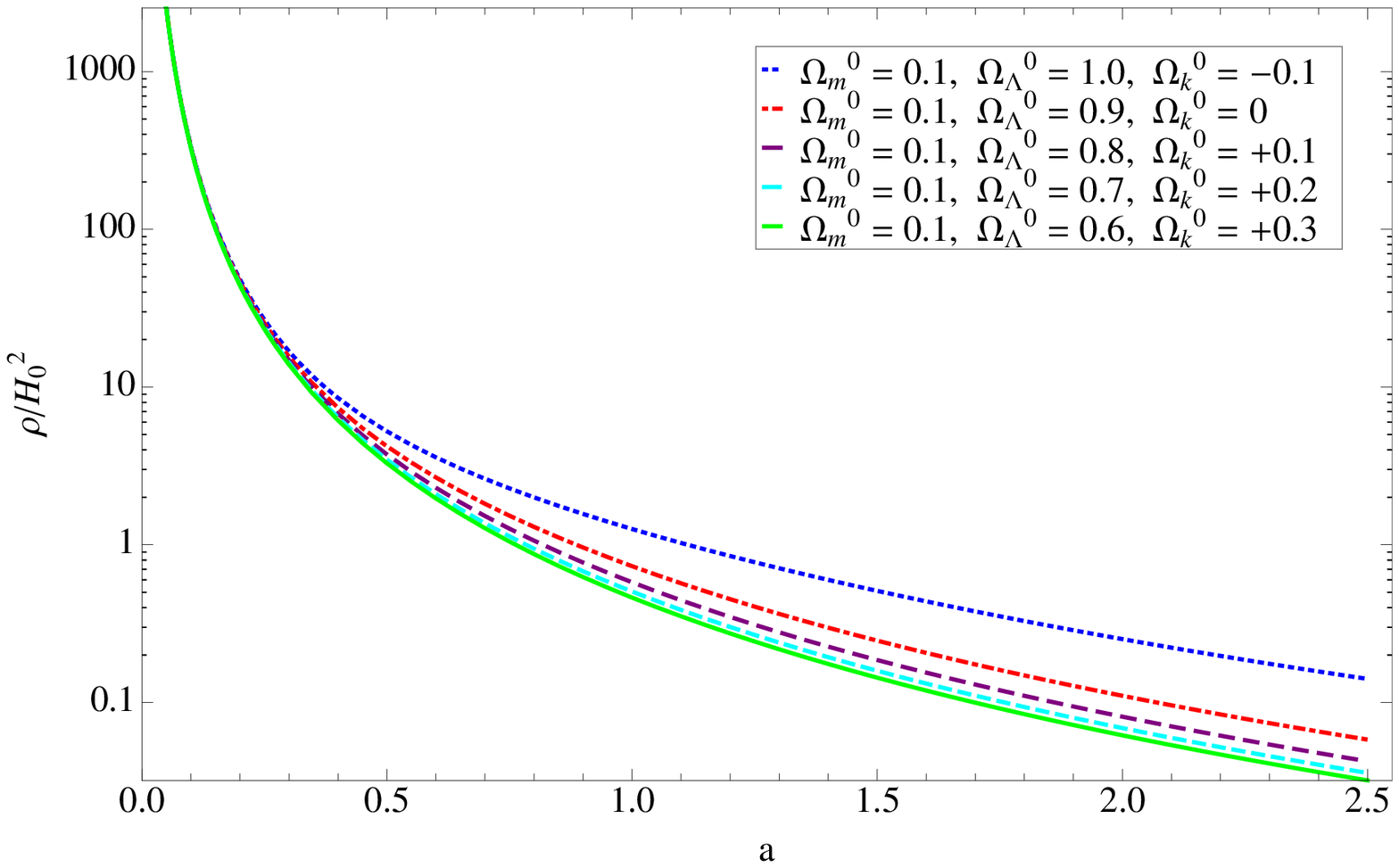}} \\
\hline
\end{tabular}
\caption{\label{fig:RhoPlot}
LEFT: Evolution of density in class II Szekeres models with no cosmological constant. As expected, all models here diverge as $a$ approaches the initial singularity, but late-time divergences within our cosmic history happen at times that depend on the combination of matter and curvature parameters. For example, the $\Omega_m^0=1$ case diverges at $a\approx1$, followed by the model with $\Omega_m^0=0.8$, and models with less matter experience density divergences in the future at later $a$ values. RIGHT: Evolution of density in Szekeres models with a cosmological constant and fixed $\Omega_m^0=0.1$. Again, all models diverge at the initial singularity, but they remain finite up to today and tend to zero as $a$ becomes large. Positively curved models ($\Omega_\Lambda^0>0.9$) have density curves lying above the negatively curved models ($\Omega_\Lambda^0<0.9$) with the flat case in between. The addition of a cosmological constant avoids the late-time divergences in the density.
}
\end{center}
\end{figure}

The left panel of Fig. \ref{fig:RhoPlot} shows $\rho/\mathbb{H}_0^2$ for class II models without a cosmological constant. As can be understood from Eq. (\ref{rhoG}), all models there diverge as $a$ approaches the initial singularity, but late-time divergences within our cosmic history depend on the values of the matter and curvature parameters. The matter-only case diverges at $a\approx0.5$, and models with less matter (larger $\Omega_k^0$) also have densities that diverge at later $a$ with smaller $\Omega_m^0$ resulting in a later onset of divergence. The right panel shows the evolution of density in Szekeres models that include a cosmological constant and have $\Omega_m^0$ set to $0.1$. The models here also all diverge at the initial singularity, but they remain finite up to today and tend to zero in the future, decreasing monotonically. Models with $\Omega_\Lambda^0>0.9$ (positively curved) have density curves lying above those with $\Omega_\Lambda^0<0.9$ (negatively curved), and the flat case falls in between. 

We now turn to the tidal gravitational field, and for that we consider the electric part of the Weyl conformal curvature tensor. Its components are defined as
\be
E_{\mu\nu}=C_{\mu\alpha\nu\beta}u^\alpha u^\beta,
\label{Etensor}
\ee
where the Weyl tensor in four dimensions has components
\be
C_{\mu\nu\alpha\beta}=R_{\mu\nu\alpha\beta}+\frac{1}{2}(g_{\mu\beta}R_{\nu\alpha}+g_{\nu\alpha}R_{\mu\beta}-g_{\mu\alpha}R_{\nu\beta}-g_{\nu\beta}R_{\nu\alpha})+\frac{R}{6}(g_{\mu\alpha}g_{\nu\beta}-g_{\mu\beta}g_{\nu\alpha}).
\label{Weyl}
\ee
The magnetic part of the Weyl tensor $H_{\mu\nu}$ is zero in the Szekeres solution, and the models therefore fall into the category of ``silent" cosmological models. The shear and the electric part of the Weyl tensor are related via the propagation equation for the shear
\be
\dot{\sigma}^\mu{\,}_\nu+\sigma^\mu{\,}_\lambda \sigma^\lambda{\,}_\nu+\frac{2}{3}\Theta{\,}\sigma^\mu{\,}_\nu-\frac{1}{3}(\delta^\mu{\,}_\nu+u^\mu\,u_\nu)  \sigma^2=-E^\mu{\,}_\nu,
\label{ShearPropag}
\ee
which reveals how $E_{\mu\nu}$, interpreted as the tidal gravitational field \cite{Ellis1971,Ellis&VanElst1998}, causes shearing in the fluid flow.

We are interested in how the tidal field scalar, the magnitude of which is given by
\be
	E^2=E^\mu{}_\nu E^\nu{}_\mu,
\ee 
and whose mixed components are \cite{typo2}
\be
	2E^{\tilde{x}}\,_{\tilde{x}}=2E^{\tilde{y}}\,_{\tilde{y}}=-E^r\,_r=\frac{2}{3}\left(\frac{\ddot{H}}{H}+2\frac{\dot{a}}{a}\frac{\dot{H}}{H}\right),
\ee
compares to the energy density at early and late cosmic times. The class I expression has the simple form
\be
	E^2=6\left(\frac{M}{a^3}\hat\delta\right)^2,
\ee
and in class II, in terms of cosmological parameters evaluated today, a small calculation gives
\be
	E^2 = \frac{3}{2}\left(\frac{\Omega_m^0}{a^2}\mathbb{H}_0^2 G\right)^2.
\ee
\begin{figure}
\begin{center}
\begin{tabular}{|c|c|}
\hline
{\includegraphics[scale=0.49,angle=0]{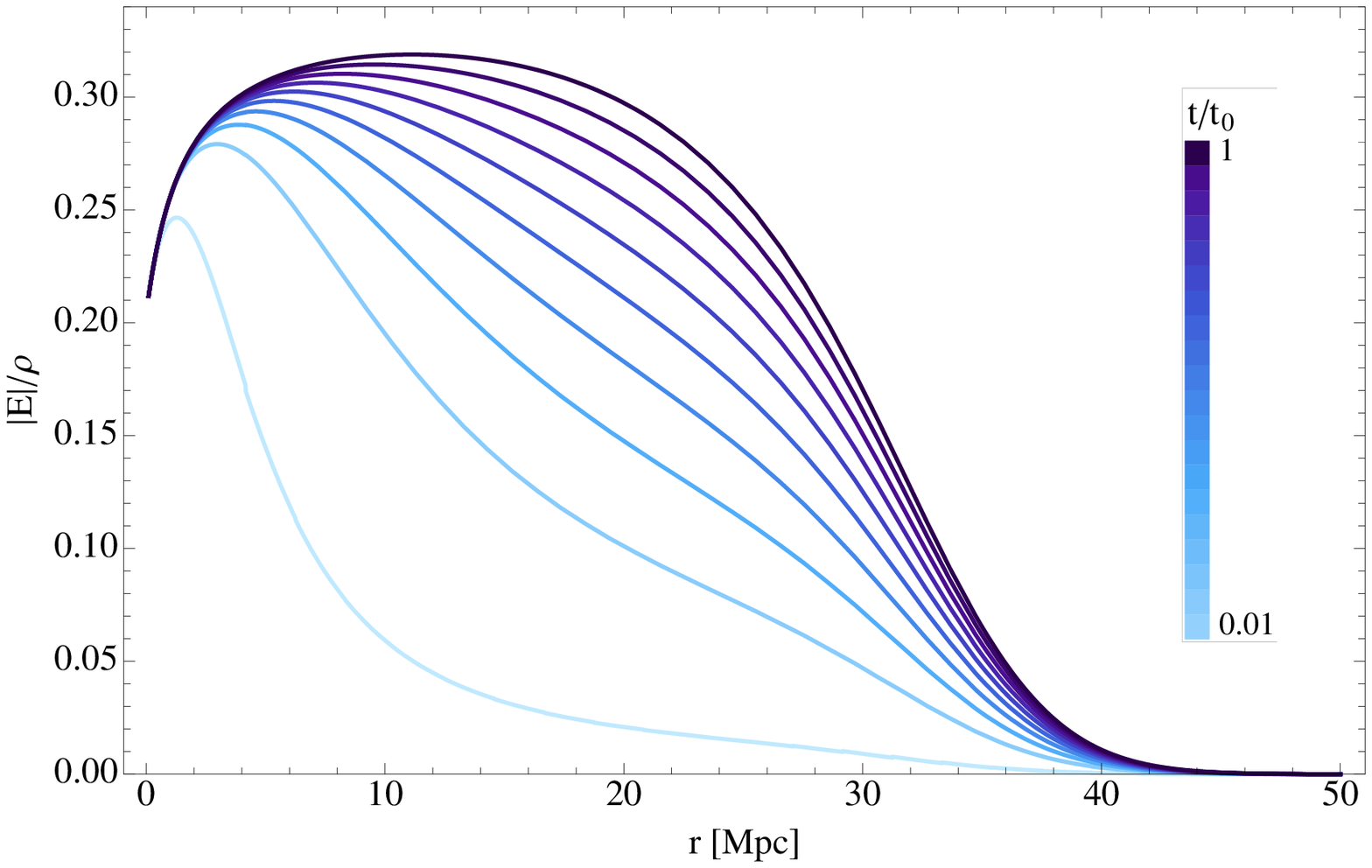}}&
{\includegraphics[scale=0.49,angle=0]{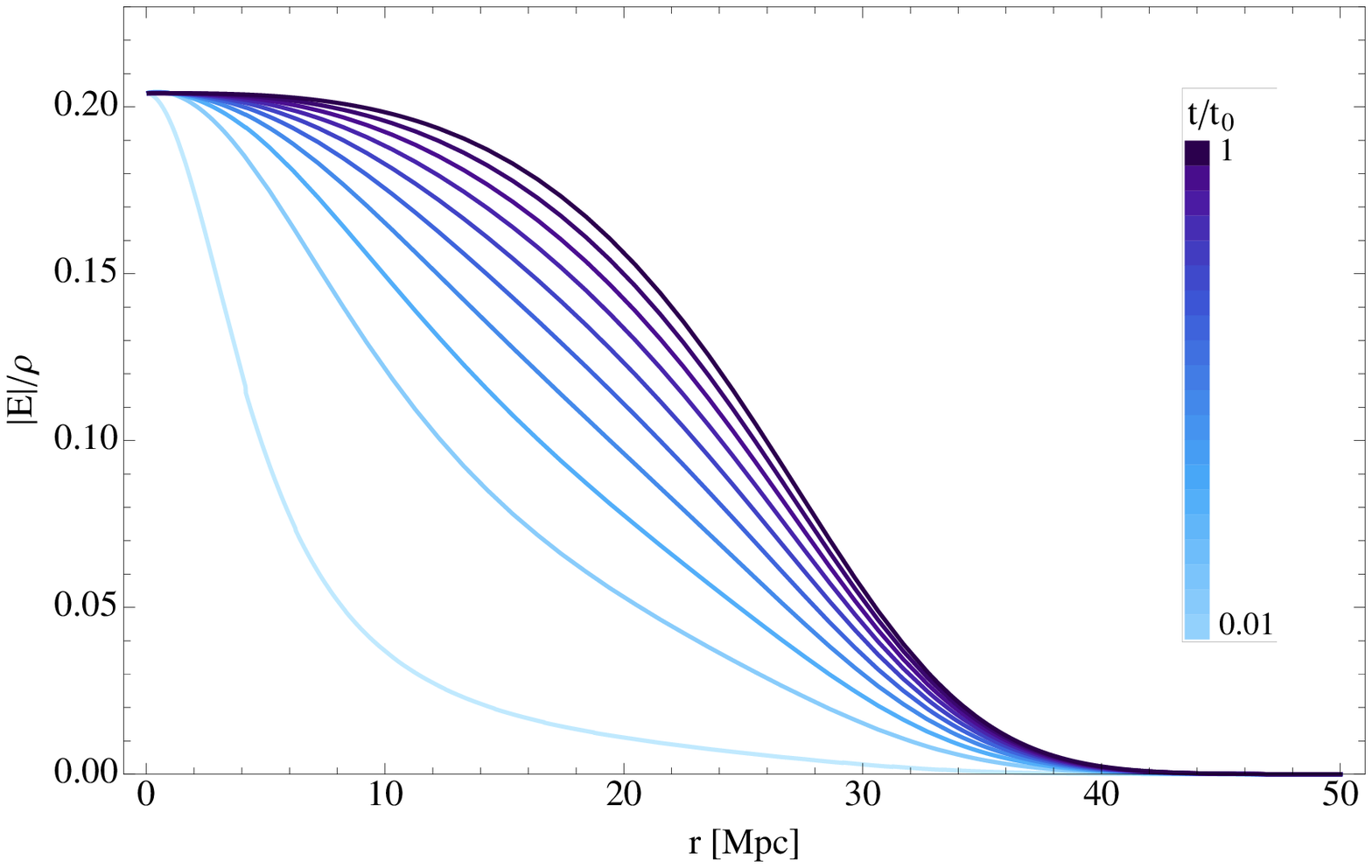}} \\
\hline  
\end{tabular}
\caption{\label{fig:classItidalE}
LEFT: Evolution of $|E|/\rho$ for the class I model along a direction that passes through the supercluster neighboring the central void. The color gradient represents $|E|/\rho$ at different times where the present time is darkest and curves get lighter toward the past. It evolves from a localized peak around $r\approx1$ Mpc in the past to a larger and significantly broader maximum around $r\approx11$ Mpc today.
RIGHT: Evolution of $|E|/\rho$ along a direction orthogonal to the supercluster. Here, the tidal field-to-density ratio is a monotonically decreasing function of $r$ that diminishes continuously toward the past. In both plots, $|E|/\rho$ approaches zero at $r=50$ Mpc, which we expect due to matching to the almost FLRW background there and that the Weyl tensor is zero in FLRW models. In addition, as adjacent curves are spaced equally in time, the tidal field was changing more quickly with respect to the density in the past than it is today. 
}
\end{center}
\end{figure}

Again, in the expressions for $\rho$ and $E^2$, the dependence on the cosmological constant comes via $a$ and $\Omega_\Lambda^0$, which is contained within $G$. Figure \ref{fig:classItidalE} shows the ratio $|E|/\rho$ along the two directions considered previously for our class I model described in Appendix C. The left panel is a radial profile through the supercluster, and the right panel is plotted for an orthogonal direction. In both cases the tidal field-to-density ratio increases continuously up to the present, but in the direction of the overdensity, the curves evolve toward wide peak around $r\approx11$ Mpc. In the other direction, the curves are monotonic decreasing functions of $r$. The matching to the almost FLRW background at $r=50$ is again clear in both panels, which we expect since the Weyl tensor is zero in FLRW models. Also, since the curves are spaced evenly in time, we see that $|E|$ was changing more rapidly compared to $\rho$ in the past than it is at present. These observations are reminiscent of the behavior of $\hat\delta$ itself, which is not surprising, because $|E|/\rho$ can be written in terms of only $\hat\delta$ as $|E|/\rho=\hat\delta/[\sqrt{6}(1+\hat\delta)]$.
\begin{figure}
\begin{center}
\begin{tabular}{|c|c|}
\hline
{\includegraphics[scale=0.49,angle=-0]{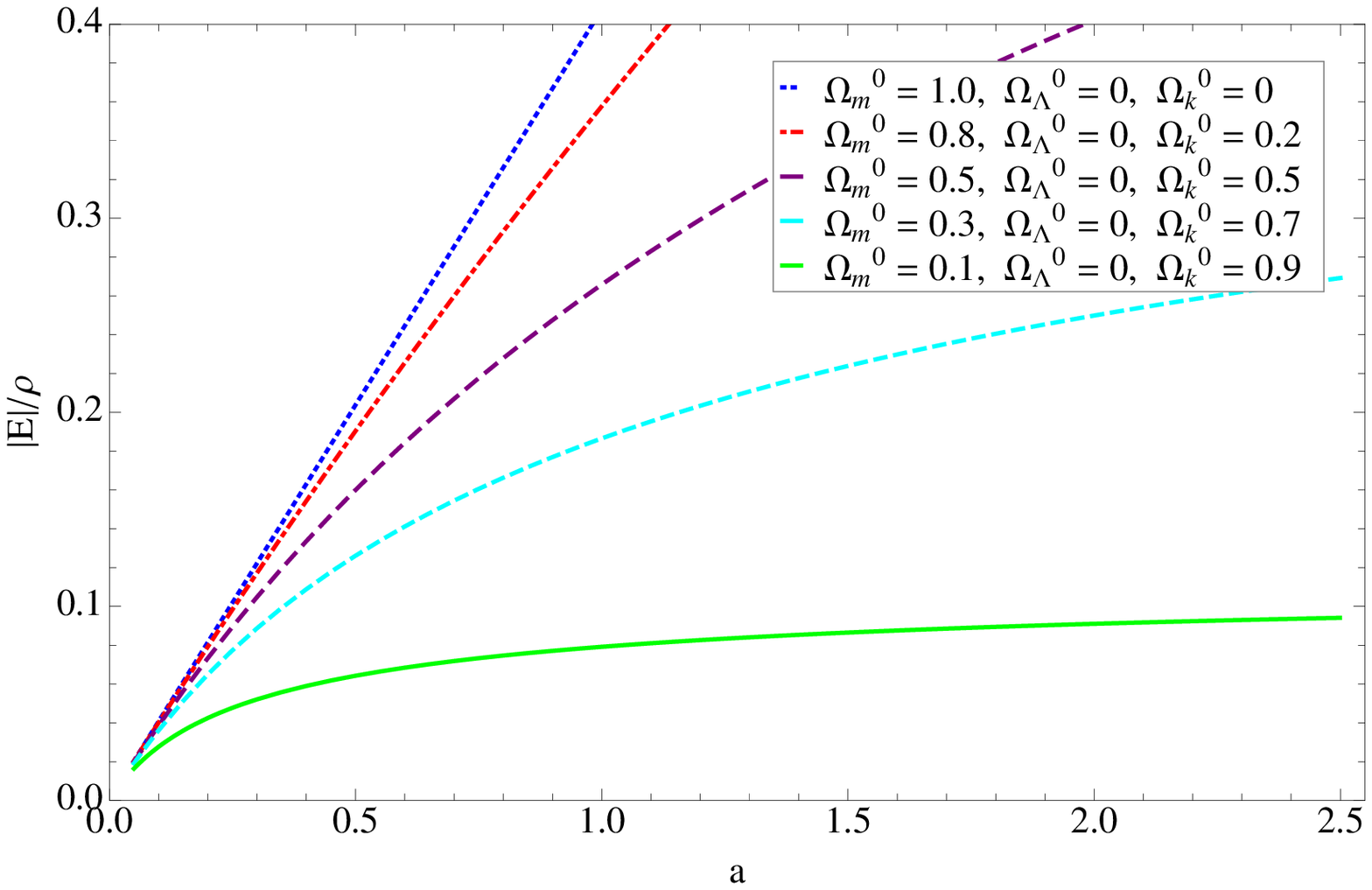}}&
{\includegraphics[scale=0.48,angle=-0]{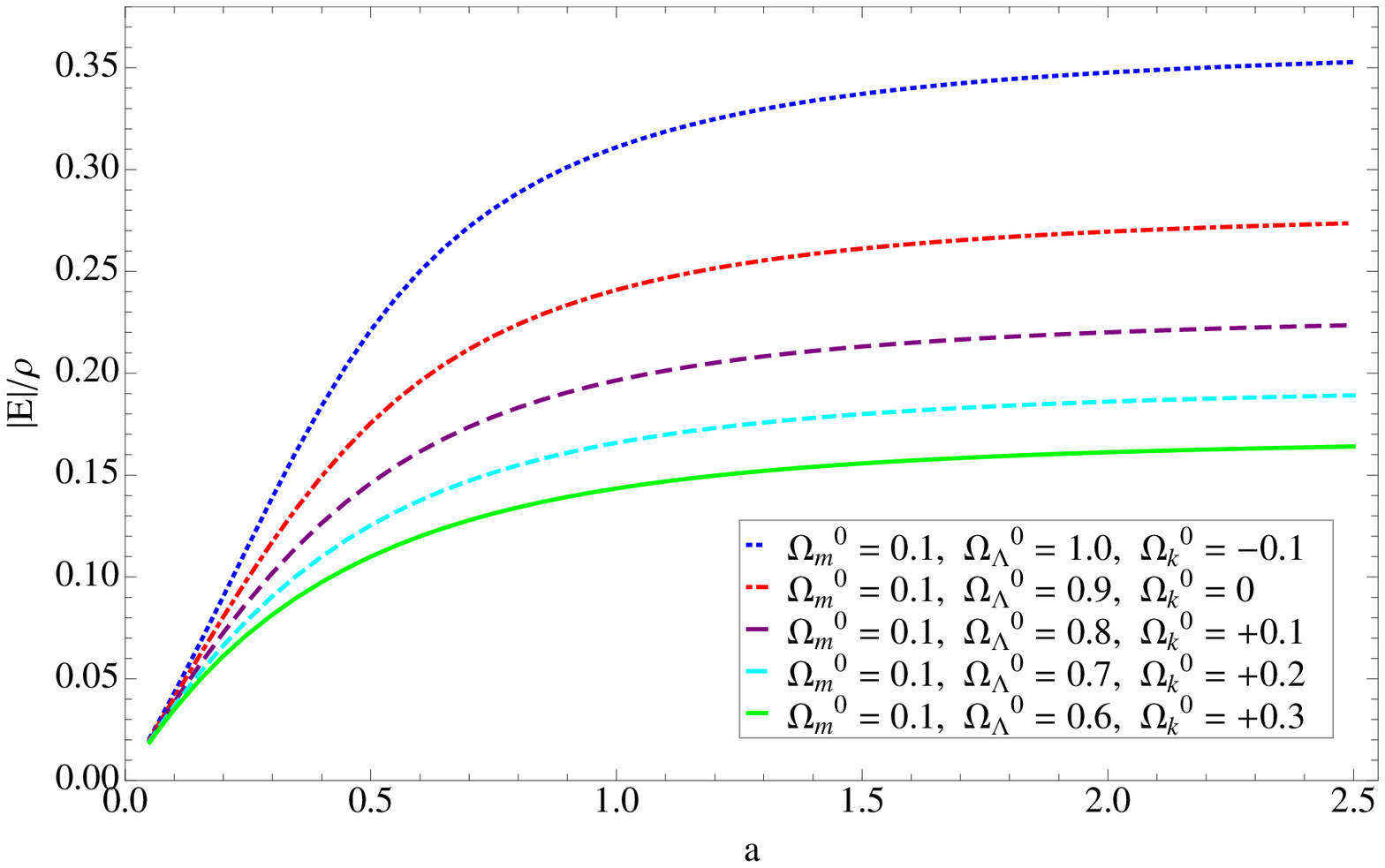}} \\
\hline
\end{tabular}
\caption{\label{fig:EoverRhoPlot}
LEFT: The ratio of the tidal gravitational field scalar to the density in class II Szekeres models with no cosmological constant. The curves all grow monotonically from zero at the initial singularity and remain finite within our cosmic history. The larger the matter content, the larger the ratio of $|E|$ to $\rho$, and the less suppression the curve experiences at later times. RIGHT: The ratio of the tidal gravitational field scalar to the density in class II models with a cosmological constant and fixed $\Omega_m^0=0.1$. As in the case without $\Omega_\Lambda$, all the curves grow monotonically from zero, but here they all also evolve almost identically after $a\approx1$. As $a$ becomes larger, $|E|$ and $\rho$ for a given set of parameters ultimately grow at the same rate so that their ratio approaches a constant. The larger $\Omega_\Lambda^0$ curves have larger $|E|/\rho$ than those with smaller $\Omega_\Lambda^0$ for the same matter content. Again, the addition of a cosmological constant removes the late-time divergences seen to the left. 
}
\end{center}
\end{figure}

We plot $|E|/\rho$ in Fig. \ref{fig:EoverRhoPlot} and compare early- and late-time divergences in class II models with $\Lambda$ to those without. On the left are models without a cosmological constant, and we see that all curves grow monotonically from zero at $a=0$ and remain finite within our cosmic history. As with $|\sigma|/\Theta$ and $\rho/\mathbb{H}_0^2$, the matter-only case is on top, and larger matter content corresponds to a larger ratio of $|E|$ to $\rho$. On the right we plot Szekeres models with a cosmological constant and fixed $\Omega_m^0=0.1$. As in the case without $\Omega_\Lambda$, all the curves grow monotonically from zero, but here they all evolve almost identically after $a\approx1$, indicating that $|E|$ and $\rho$ grow proportionally at larger $a$. Again, similar to $|\sigma|/\Theta$ and $\rho/\mathbb{H}_0^2$, larger $\Omega_\Lambda^0$ curves overall have larger $|E|/\rho$ than those with smaller $\Omega_\Lambda^0$ for the same matter content, with the flat ($\Omega_\Lambda^0=0.9$) case falling in between. The important point is that for all the models plotted here, the addition of a cosmological constant removes any late-time divergences that were present without it. 

Furthermore, since the cosmological constant will continue to dominate as time increases, one expects a future evolution free from singularities for the models of interest. The Szekeres solutions can have shell-crossing singularities when the metric function $H$ vanishes. This is clear from Eq. (\ref{density}) for the matter density. We have verified that for the models of interest explored above with $\Omega_m^0\lesssim0.1$, the growth rate and the density do not become singular in the future. $G$ approaches zero as $a$ becomes large, and since the scale factor grows without bound in these models, the density goes to zero along with $G$.

%
\section{Class II Comparison to Current Growth Data}\label{growthFactor}
In the preceding sections, we have demonstrated that the Szekeres class II models are capable of reproducing the required features of the growth history of large-scale structure using a nonzero cosmological constant $\Lambda$ to mimic the $\Lambda$CDM concordance model. This includes both a late-time suppression of growth due to $\Lambda$ and significantly stronger growth at early times than is possible with linearly perturbed $\Lambda$CDM using only a baryonic matter component $\Omega_b^0=0.04$ consistent with constraints from Big Bang nucleosynthesis (BBN). While we were able to explore an initial parameter space for the Szekeres models by qualitative comparison to $\Lambda$CDM, we will now compare the models directly to currently available growth data. 

The following discussion only applies to class II models. We do not attempt to compare class I models to available growth data, as this requires a full consideration of the cosmological parameters as well as the available data as functions of $r$. There is not enough growth data points at the present time in order to attempt such $r$-modelization and we leave this for future work while specializing the current comparison to class II. 

\begin{table}
\center
\begin{tabular}{ c c c }
\hline
z~~ & $f_{\textrm{obs}}$~ & Ref. \\
\hline
0.15~~ & $0.49\pm 0.10$~ & \cite{19,20} \\
0.20~~ & $0.60\pm 0.10$~ & \cite{27} \\
0.35~~ & $0.70\pm 0.18$~ & \cite{21} \\
0.40~~ & $0.70\pm 0.07$~ & \cite{27} \\
0.55~~ & $0.75\pm 0.18$~ & \cite{22} \\
0.60~~ & $0.73\pm 0.07$~ & \cite{27} \\
0.80~~ & $0.70\pm 0.08$~ & \cite{27} \\
0.77~~ & $0.91\pm 0.36$~ & \cite{19} \\
1.40~~ & $0.90\pm 0.24$~ & \cite{23} \\
2.42~~ & $0.74\pm 0.24$~ & \cite{25} \\
2.60~~ & $0.99\pm 1.16$~ & \cite{26} \\
2.80~~ & $1.13\pm 1.07$~ & \cite{26} \\
3.00~~ & $1.66\pm 1.35$~ & \cite{26} \\
3.00~~ & $1.46\pm 0.29$~ & \cite{24} \\
3.20~~ & $1.43\pm 1.34$~ & \cite{26} \\
3.40~~ & $1.30\pm 1.50$~ & \cite{26} \\
\hline
\end{tabular}
\caption{Summary of observational data used to constrain the cosmological parameters in the Szekeres class II and $\Lambda$CDM models through the growth factor. Data is taken from the compilation of \cite{28} and \cite{27}, with original references for each value listed in the right column. The first nine values are derived from measurements of redshift space distortions, while the last seven are from Lyman-$\alpha$ measurements. Where necessary, we have given the mean redshift of the $f$ measurements when originally reported for a range of redshifts.}
\label{table1}
\end{table}

We will use measurements of the growth factor 
\be
f=\mathrm{d}\ln\delta/\mathrm{d}\ln a,
\label{fdef}
\ee
 which has been constrained over a wide range of redshifts. The values of $f$ that we compare to are listed in Table \ref{table1} and come from measurements of both redshift space distortions and the Lyman-$\alpha$ forest. The growth factor is derived from either the redshift distortion parameter $\beta=f/b$ (where b is the deterministic galaxy bias) or from the amplitudes of power spectra from Lyman-$\alpha$ forest data. Certain assumptions based on $\Lambda$CDM are necessary to obtain this growth data for $f$ and thus may impact a direct comparison of the data to the growth resulting from modeling the evolution of structure in a Szekeres universe. However, we explore here the formalism of such a comparison and provide some first insight into the Szekeres models' ability to fit real growth data.
%
\subsection{Growth factor and growth index in the Szekeres models}
%
In $\Lambda$CDM, the growth factor is written from the growth equation as
\be
f'+\left(2+\frac{\dot{\mathrm{H}}}{\,\,\mathrm{H}^2}\right)f+f^2-\frac{3}{2}\Omega_{m}=0,\label{eq:flcdm}
\ee
where $\mathrm{H}$ here is the $\Lambda$CDM Hubble parameter; see for example \cite{linder,polarski,Gong09,Mortonson,linder07,gannouji,PaperI,28}.
\begin{table}
\center
\begin{tabular}{ l c | c c c c | c c c c }
\hline
 & & \multicolumn{4}{c |}{Without priors} & \multicolumn{4}{ c}{With priors}\\
  & L$\alpha$? & $\Omega_m^0$ & $\Omega_{\Lambda}^0$ & $\Omega_k^0$ & ($\chi^2$) & $\Omega_m^0$ & $\Omega_{\Lambda}^0$ & $\Omega_k^0$ & ($\chi^2$) \\
\hline
\multirow{2}{*}{Szekeres} & No & 0.12 & 0.59 & 0.29 & (0.22) & 0.05 & 0.98 & -0.03 & (0.62) \\
 & Yes & 0.11 & 0.69 & 0.20 & (0.39) & 0.05 & 0.98 & -0.03 & (0.63) \\
\hline
\multirow{2}{*}{$\Lambda$CDM} & No & 0.29 & 0.56 & 0.15 & (0.21) & 0.27 & 0.73 & 0.00 & (0.32) \\
 & Yes & 0.26 & 0.69 & 0.05 & (0.39) & 0.27 & 0.73 & 0.00 & (0.43) \\
\hline
\end{tabular}\caption{Best fit parameters for Szekeres and $\Lambda$CDM models. We use the $\chi^2$ minimization described in Eq. (\ref{eq:chi}) in order to choose the best cosmological parameters that fit $f$ to the observed growth factor as given in Table \ref{table1}. We present results both including and omitting available Lyman-$\alpha$ data. We also present results with priors placed on the Szekeres model, such that $0.039\le\Omega_b^0\le 0.049$ from BBN (with $\Omega_{dm}^0=0$). For comparison, we include the $\chi^2$ for each best fit along with the $\chi^2$ for standard $\Lambda$CDM parameters. Values are rounded to the nearest percent. The Szekeres model with no priors fits the growth data with approximately the same $\chi^2$ as that of the $\Lambda$CDM model.}
\label{table2}
\end{table}

In a similar way to previous work done for $\Lambda$CDM, we can derive the growth factor for the Szekeres models. From the definition of $f$ in Eq. (\ref{fdef}), we substitute 
\be
	\delta'=f\frac{\delta}{a}
\ee
and
\be
	\delta''=\left(f^2-f+f'\right)\frac{\delta}{a^2}
\ee
into Eq. (\ref{growth_a2_classI}). Simplifying the resulting expression gives
\be
f'+\left(1+\Omega_{\Lambda}-\frac{1}{2}\Omega_{m}\right)f+\left(1-\frac{2}{1+X^{-1}}\right)f^2-\frac{3}{2}\left(1+X\right)\Omega_{m}=0,
\ee
where
\be
X=\delta=\delta(a=0)\exp\left(\int f\,\mathrm{d}\ln a\right).
\ee

Unlike for $\Lambda$CDM, the higher order terms in $\delta$ of Eq. (\ref{growth_a2_classI}) in the Szekeres models force us to keep the integral form for $X$. One could attempt to integrate this numerically as has been done in the past. Alternatively, as in our case, we can simply observe that $f$ is related to the growth rate $G$ in Eq. (\ref{growth_G_classII}) as
\be
f=1+a\frac{G'}{G},
\ee
where $G'$ and $G$, and thus $f$, are all functions of $a$, $\Omega_m^0$, and $\Omega_{\Lambda}^0$, and we require $\Omega_k^0=1-\Omega_m^0-\Omega_{\Lambda}^0$. Thus the function $f$ can be calculated from $G$ and $G'$, which have already been integrated through the Runge-Kutta method used in the first sections of this work. In the next subsection, we compare the function $f$ in the Szekeres models directly to measured values of $f$. 

Also important in studies of the growth of structure is the growth index $\gamma$. The growth factor $f$ was shown to be well approximated by \cite{Peebles1980,Fry,Lightman,wang}
\be
f=\Omega_m^{\gamma}.
\ee
More recently, the growth index has been demonstrated to be a useful parametrization of $f$, capable of distinguishing between $\Lambda$CDM and other alternative dark energy and modified gravity models, having different characteristic values in each \cite{linder,polarski,Gong09,Mortonson,linder07,gannouji,PaperI,28}. It can be written in $\Lambda$CDM with a dark energy equation of state parametrized by $w$ as an expansion to lowest order in $1-\Omega_m$ and $\Omega_k$ \cite{Gong09}
\be
\gamma=\frac{3(w-1)(1-\Omega_m)-(3w+1)\Omega_k}{(6w-5)(1-\Omega_m)-2(3w+1)\Omega_k}.\label{eq:gamma}
\ee
For early times ($\Omega_m=1$) and zero curvature, we find $\gamma=6/11$. This result is nearly independent of the choice of $\Omega_m$, $\Omega_{\Lambda}$, and $\Omega_k$, with the value of $\gamma$ changing only at the percent level. However, it has been shown that the original ansatz $\ln f=\gamma\ln\Omega_m$ must be altered in some cases, such as for a curved space. It is thus not immediately clear that we can apply directly a relationship like $f=\Omega_m^{\gamma}$ for some $\gamma$ in a general Szekeres space, and the derivation of an analytic form for such a $\gamma$ is not trivial if possible at all. We will explore whether a $\gamma$ can characterize the Szekeres models in the following section.

\begin{figure}
\includegraphics[angle=270,scale=0.49]{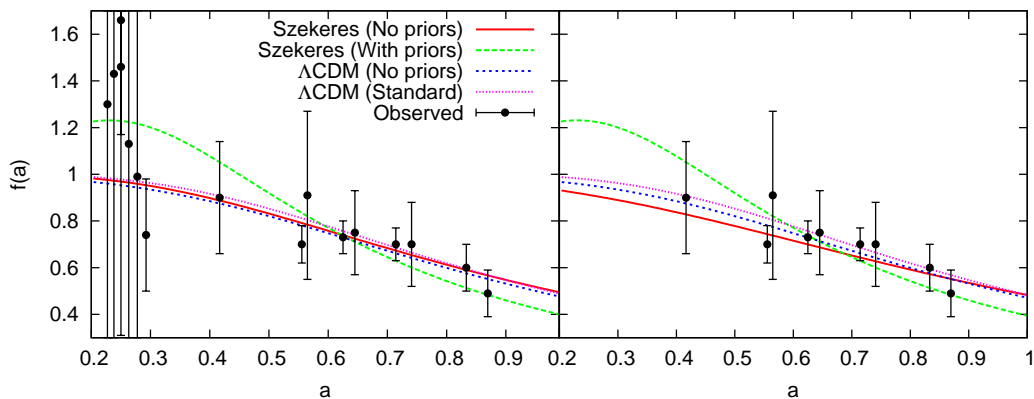}
\vskip.5in
\caption{Plots of the best fit growth factor $f$ to observations for Szekeres and $\Lambda$CDM models. Best fit parameter values are determined by $\chi^2$ minimization with and without priors. The main result plotted here is that the Szekeres model with no priors fits the growth data with $\chi^2$ almost indistinguishable from those of the $\Lambda$CDM. The best fit Szekeres model requires less matter density and more spatial curvature. For further exploration, priors on the Szekeres model are applied and consist in constraining the total matter density to be that of only baryons, with values as limited by the BBN (i.e. $0.039\le\Omega_b^0\le 0.049$). Also, the priors when imposed on $\Lambda$CDM consist in fixing the density parameters to those of the concordance model with  $\Omega_m^0=0.27$ and $\Omega_{\Lambda}^0=0.73$. 
The plots are shown with growth factor data both including (left panel) and excluding (right panel) available Lyman-$\alpha$ data for the parameter sets described in Table \ref{table2}. 
}\label{fig:fcomp}
\end{figure}
\subsection{Fitting and results}

We use the measured values of $f$ given in Table \ref{table1} to constrain the best fit cosmological parameters $\Omega_m^0$, $\Omega_{\Lambda}^0$, and $\Omega_k^0$ using a $\chi^2$ minimization. We define the $\chi^2$ for $N$ data points as
\be
\chi^2=\frac{1}{N}\sum_{i=1}^{N}\frac{f^2(a_i)-f_{\textrm{obs}}^2(a_i)}{\sigma^2(a_i)},\label{eq:chi}
\ee
where $\sigma(a_i)$ is the absolute error in the measurement at $a_i$. We construct the $\chi^2$ in this way so as to be able to directly compare results that use a different number of data points. The $\chi^2$ is then one when the average difference in $f$ from observations is equal to the average error. 

We first calculate the $\chi^2$ with no priors on the values of the cosmological parameters for both the Szekeres models and $\Lambda$CDM, and then impose for the Szekeres models the requirement that $0.039\le\Omega_b^0\le 0.049$ from BBN \cite{wmap7}, assuming that there is no dark matter component ($\Omega_{dm}^0=0$). For comparison, we also calculate the $\chi^2$ for the concordance $\Lambda$CDM cosmology with values $\Omega_m^0=0.27$ and $\Omega_{\Lambda}^0=0.73$. Finally, we determine the parameters for the Szekeres models which best fit the growth in the standard $\Lambda$CDM cosmology. We summarize these results in Table \ref{table2} and compare the resulting $f(a)$ to $f_{\textrm{obs}}(a_i)$ in Fig. \ref{fig:fcomp}.

We find that there is negligible difference in fitting power ($\chi^2$) between our best fit Szekeres parameters with no priors and those for $\Lambda$CDM, both including and excluding Lyman-$\alpha$ data. 
In general, we find that the Szekeres models fit the measured $f$ with comparable $\Omega_\Lambda^0$ to that of the $\Lambda$CDM model, but with a lower matter content and a larger spatial curvature.
In order to explore the Szekeres fitting result further, we include the priors from BBN that $0.039\le\Omega_b^0\le 0.049$ and set $\Omega_{dm}^0=0$. We find in this case that the Szekeres models perform much worse in fitting the measurements, with almost a factor of three increase in $\chi^2$ when excluding the Lyman-$\alpha$ data. This indicates that while the Szekeres requires smaller values for the dark matter density, it still needs some in order to have a good fit to the growth data. As a simple check, we find that the concordance $\Lambda$CDM parameters are still a good fit to the growth data but with non-negligible increases in $\chi^2$ when compared to the best fit parameter choices. 
For further comparison to the $\Lambda$CDM model, we find the Szekeres models most closely mimic the growth of the standard $\Lambda$CDM model with parameters $(\Omega_m^0,\Omega_\Lambda^0,\Omega_k^0)=(0.11,0.71,0.18)$ and $\chi^2=2.4\times10^{-5}$.

The information in Table \ref{table2} is represented visually in Fig. \ref{fig:fcomp}, where the various $f$ fits are generally very consistent and fit well to the data, except for the Szekeres fit with the priors listed above, which clearly fits the data very poorly both by visual inspection and $\chi^2$ value. We also note that in the right panel, where we exclude Lyman-$\alpha$ data, the best fit Szekeres $f$ with no priors is able to fit all data points, while the standard $\Lambda$CDM falls just outside the error bar on the data point at $a=0.56$ ($z=0.8$). 

Our finding of a nonzero curvature using the Szekeres models is consistent with other studies of inhomogeneous models and averaging studies, which have shown that using a more general, inhomogeneous cosmology strongly impacts the determination of cosmological parameters. For example, by introducing inhomogeneities to explore changes in the distance to the cosmic microwave background, \cite{bolcmbdist} found that parameter determinations were strongly affected, with a positive curvature and $\Omega_{\Lambda}^0\approx 0.8-0.9$ preferred. This is similar to our determination of the parameters for the Szekeres models in the case where we require only baryonic matter to contribute to the matter density and find a small positive curvature and large $\Omega_{\Lambda}^0$. Similarly, \cite{buchert} explored the effects of averaging inhomogeneities on the measurement of cosmological parameters and found that without prior assumptions on $\Lambda$, a substantial negative curvature was favored. This is also consistent with our best fit determination of the Szekeres models, which contains a significant negative curvature, as well as lower matter density than required in $\Lambda$CDM. Further, \cite{coley} found that the averaging will in general affect the Friedmann equation by introducing a curvaturelike term. Thus the measurement of an FLRW model with negligible curvature, which is considered to be associated with the averaged background of a general Szekeres model, may indicate that a nonzero curvature in the true Szekeres model is appropriate.

\begin{figure}[t]
\includegraphics[angle=270,scale=0.35]{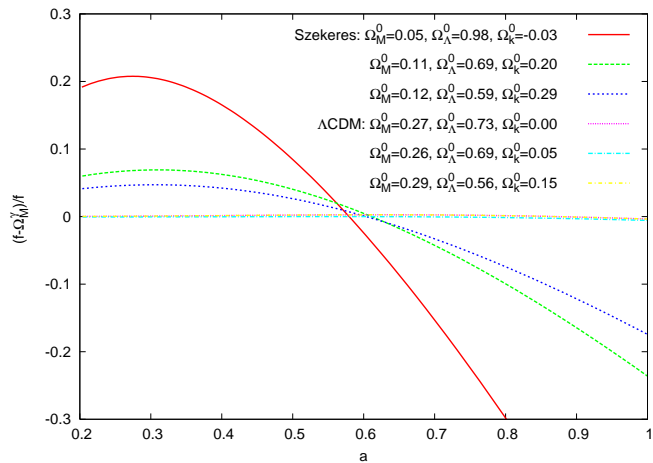}%
\caption{The relative error $(f-\Omega_m^{\gamma})/f$ of the three unique parameter fits in Table \ref{table2} for both Szekeres and $\Lambda$CDM models. The relative errors of the three $\Lambda$CDM parameter sets overlap and are nearly zero even at late times, while there is significant error in the three Szekeres parameter sets, particularly at late times.}\label{fig:gcomp}
\end{figure}

In addition to fitting the growth factor, we also attempt to fit a growth index. To check whether we can apply the ansatz $\ln f=\gamma\ln\Omega_m$ to the Szekeres models, we perform a numerical $\chi^2$ minimization to find the best fit $\gamma$ for the resulting parameters in Table \ref{table2}. As a check that this fitting method performs well, we first apply it to the $\Lambda$CDM parameter sets, finding as expected that $\gamma\approx 0.55$. We find, however, that there does not exist a unique $\gamma$ for the Szekeres models, but instead that $\gamma$ varies with different $\Omega_m$, and we find very poor $\chi^2$ values for these fits. Further, there exists only a small subset of parameter choices for which we can fit a $\gamma$ at all, indicating that the usual ansatz used for FLRW models fail for the Szekeres models. This is demonstrated in Fig. \ref{fig:gcomp}, which shows a plot of the relative error $(f-\Omega_m^{\gamma})/f$ of the three unique parameter fits in Table \ref{table2} for both Szekeres and $\Lambda$CDM models. While the relative error in the $f=\Omega_m^{\gamma}$ relationship for $\Lambda$CDM overlaps for the three parameter sets and is nearly zero even at late times, there is significant error in the three Szekeres parameter sets, particularly at late times showing that the usual ansatz cannot be used here and direct comparison to the growth factor data is required.  

We conclude then that the Szekeres models are able to exactly mimic a standard $\Lambda$CDM growth history characterized by $f$, and to fit observed values of the growth factor with comparable accuracy to the $\Lambda$CDM model. The Szekeres models require comparable $\Omega_{\Lambda}^0$ to $\Lambda$CDM, but significantly less dark matter, which is offset by a large negative curvature. Our parameter determinations are roughly consistent with previous studies of inhomogeneous models and averaging. We also find that the growth index parameterization fails for the Szekeres models, where a single-value $\gamma$ is unable to fit $f$. However, any conclusions based on comparisons of the Szekeres models to observations, and in particular the determination of cosmological parameters, should be made with caution. The measured values of $f$ are based on assumptions of $\Lambda$CDM, which do not hold in a more general Szekeres model. The cosmological parameters described in the preceding sections are defined in the same way as their $\Lambda$CDM counterparts and should represent the same fractional energy contents in the universe. However, the energy densities do not have strictly the same connections to physical quantities in both cases. For example, the Hubble parameter in FLRW is proportional to the expansion scalar $\Theta$, but in Szekeres there is also a dependence on the metric function $H$ (see Sec. \ref{scalars}).  Thus, a more careful consideration is required when comparing the Szekeres parameters directly to $\Lambda$CDM.

%
\section{Conclusion}\label{conclusion}
%
The standard approach to studying large-scale cosmological structure evolution is to apply linear perturbations to a smooth FLRW metric. However, in order to go beyond the linear perturbation framework, it is worth exploring alternative exact frameworks. Such exact frameworks can include linear and nonlinear effects from for example large-scale superstructures in the lumpy universe that we observe. In this work, we have studied the growth of large-scale structure in the Szekeres inhomogeneous cosmological models in the presence of a cosmological constant and allowing for nonzero spatial curvature. Using the Goode and Wainwright formulation of the solution, we have identified a density contrast for both classes ($\hat\delta$ in class I and $\delta$ in class II) in terms of the metric functions $F$ and $H$. The $F$ function obeys a second-order time differential equation like the density contrast in FLRW, and this allows us to write evolution equations for $\hat\delta$ and $\delta$ that have linear terms formally identical to the linearly perturbed FLRW equation. However, our equations also contain nonlinear terms. Most importantly, unlike linear perturbations, the equations here are derived in an exact framework, meaning $\hat\delta$ and $\delta$ are not constrained to be smaller than $1$.

To study structure growth in class I, we have constructed (today) a simple nonsymmetric and inhomogeneous region that physically represents a central void and neighboring supercluster surrounded by an asymptotically smooth background. The model matches to an almost FLRW model with standard $\Lambda$CDM parameters at $r=50$ Mpc. By numerically solving the growth equation for $\hat\delta$, we have evolved the structure backward in time to see the region smooth out, i.e. become more homogeneous, with both the overdensity and the void shrinking continuously. The class I dependence on the coordinate $r$ of the metric functions allows us to model such a single structure, while for class II such an implantation is possible via the metric functions $\beta_{\pm}$ since $M$ is constant and $a=a(t)$. 

The class II models, as we have set them up, are described by their input cosmological parameters today and describe exact deviations from an underlying FLRW model (obtained by letting $\delta=0$). For class I models, these input cosmological parameters today can be understood to apply at some relevant $r$, and we restrict ourselves to the independently evolving spatial section defined by that constant $r$. For both classes, we find a growth up to several times stronger than that of the  $\Lambda$CDM model. For example, we find that for flat class II models with $\Omega_m^0$ and $\Omega_\Lambda^0$ values comparable to those of $\Lambda$CDM, Szekeres models experience significantly stronger growth than their linearly perturbed FLRW counterparts. The flat model with $\Omega_m^0=0.04$ provides a qualitatively consistent growth history that, if the matter content is interpreted to be solely due to baryons, appears not to require dark matter in order to achieve a similar growth rate to $\Lambda$CDM today. This value of $\Omega_m^0$ agrees with the value of $\Omega_b^0$ as predicted by big bang nucleosynthesis. Flat and curved class II models with nonzero $\Omega_\Lambda^0$ experience suppression of the growth rate after the onset of $\Lambda$ dominance, as expected. 

We have also derived expressions for kinematic quantities in both classes, including the shear scalar, expansion scalar, and electric part of the Weyl tensor, which represents the tidal gravitational field. We explored the behaviors of $|\sigma|/\Theta$, $\rho$, and $|E|/\rho$ for the class I model along directions through the supercluster and orthogonal to it, as well as for class II models with various combinations of cosmological parameters. These quantities do not experience singularities in the class I model, and in class II, the inclusion of $\Lambda$ is found to remove divergences that were present without it.

To investigate the ability of Szekeres models to match observations, we derived the evolution equation for the growth factor $f$ in class II and also expressed it directly in terms of $G$ and its derivative, the latter being more convenient for our purposes. We used a $\chi^2$ minimization to obtain best fit parameters from redshift space distortions and Lyman-$\alpha$ forest data for Szekeres and $\Lambda$CDM models both with and without priors. For Szekeres, the parameters $(\Omega_m^0,\Omega_\Lambda^0,\Omega_k^0) = (0.12,0.59,0.29)$, obtained without priors and excluding Lyman-$\alpha$, achieve the smallest $\chi^2$. Further, the growth factor for these Szekeres parameters is able to fit all the data, while the corresponding $\Lambda$CDM curve misses the error bound on one point. With no priors imposed, we find that the Szekeres models achieve the same $\chi^2$ as the $\Lambda$CDM model but require different combinations of the cosmological parameters. We have also determined that the parameterization of $f$ by the growth index $\gamma$ does not apply to the Szekeres models in a simple way. For example, the usual $\Omega_m^\gamma$ ansatz does not work, and it is necessary to directly fit $f$.

With these results, we find that the Szekeres inhomogeneous cosmological models can provide a framework to analyze the growth of large-scale structure using exact equations not limited to the linear regime or first-order terms in the density contrast. Szekeres models subjected to current growth data can provide competitive fits compared to the concordance $\Lambda$CDM perturbative scheme, but they require different values for the matter density and spatial curvature. Our results in this work will be useful in developing a complete framework based on inhomogeneous cosmological models in which current and future observations can be analyzed and interpreted.
%
\acknowledgments
%
We thank J. Dossett for useful discussions and for reading the manuscript. M.I. acknowledges that this material is based upon work supported in part by NASA under Grant No. NNX09AJ55G and by the Department of Energy (DOE) under Grant No. DE-FG02-10ER41310 and that part of the calculations for this work have been performed on the Cosmology Computer Cluster funded by the Hoblitzelle Foundation. A.P. and M.T. acknowledge that this work was supported in part by the NASA/TSGC graduate fellowship program.
%
%
\appendix
%
\section{Evolution of Szekeres Models in the GW Representation}
We describe the time and space dependencies of Szekeres models in the Goode and Wainwright (GW) representation \cite{Goode&Wainwright1982b}. The GW metric has the form
\be
	ds^2= -\mathrm{d}t^2+a^2\left[e^{2\nu}(\mathrm{d}\tilde{x}^2+\mathrm{d}\tilde{y}^2)+H^2W^2 \mathrm{d}r^2\right].
\ee
%
\subsubsection{Time dependence}
%
The solutions to the equations governing the time evolution of Szekeres models have the same form for both classes. When $\Lambda$ is nonzero, the general solution of the Friedmann equation [Eq. (\ref{Friedmann})] involves elliptic functions and can be found in, for example, \cite{Barrow&SS1984}. We give here the original forms with $\Lambda=0$ from the Goode and Wainwright paper, which are written in parametric from with parameter $\eta$:
\be
	a=M \frac{dh}{d\eta},\qquad{\rm{with}}\quad t-T=Mh,
\label{param}
\ee
where
\be
  h(\eta) = \left\{ 
  \begin{array}{l l}
    \eta-\sin\eta,  & \quad k=+1\\
    \sinh\eta-\eta, & \quad k=-1\\
    \eta^3/6,       & \quad k=0.\\
  \end{array} \right.
\label{h}
\ee
$\eta$ varies over $0<\eta<2\pi$ for $k=+1$ and $0<\eta<\infty$ for $k=0,-1$. By Eqs. (\ref{param}) and (\ref{h}), we see that the scaling function $a$ has the same time dependence as that of an FLRW dust model. As was pointed out in Sec. \ref{GWform}, a condition to have a physically meaningful solution is that $M > 0$ \cite{Goode&Wainwright1982b}, and when $k=0,-1$ we take $\dot{a}>0$, which corresponds to an expanding universe.

The two independent solutions to the second-order differential equation in $F$ [Eq. (\ref{Ray1})], are
\be
  f_{+} = \left\{ 
  \begin{array}{l l}
    (6M/a)\,[1-(\eta/2)\cot(\eta/2)]-1, & \quad k=+1\\
    (6M/a)\,[1-(\eta/2)\coth(\eta/2)]+1, & \quad k=-1\\
    \eta^2/10, & \quad k=0\\
  \end{array} \right.\label{fplus}
\ee
and
\be
  f_{-} = \left\{ 
  \begin{array}{l l}
    (6M/a)\,\cot(\eta/2), & \quad k=+1\\
    (6M/a)\,\coth(\eta/2), & \quad k=-1\\
    24/\eta^3, & \quad k=0.\\
  \end{array} \right.\label{fminus}
\ee
\subsubsection{Space dependence}
Unlike the time dependence, the space dependence is different for each class.

{\begin{center}{class I:\qquad$a=a(t,r),\;\;a,_r\ne0,\;\;f_{\pm}=f_{\pm}(t,r),\;\;T=T(r),\;\;M=M(r)$,}\end{center}}
with
\be
e^\nu=f(r)[a(r)(\tilde{x}^2+\tilde{y}^2)+2b(r)\tilde{x}+2c(r)\tilde{y}+d(r)]^{-1},
\ee 
\bea
\epsilon/4&=& a(r)d(r)-b^2(r)-c^2(r),\qquad\;\;\; \epsilon=0,\pm 1, \nonumber \\
A&=&f\nu,_{r}-k\beta_+,\, \qquad\;\;\;W^2=(\epsilon-kf^2)^{-1},\nonumber \\
\beta_+&=&-kfM,_{r}/(3M),\, \quad \beta_-=fT,_{r}/(6M),
\label{classIbeta}
\eea
and where a comma denotes differentiation with respect to the variable that follows. The functions $f(r)$, $a(r)$, $b(r)$, $c(r)$, $d(r)$, and $T(r)$ are smooth but otherwise arbitrary. The function $a(r)$ is distinct from the generalized scale factor $a(t,r)$.

{\begin{center}{class II:\qquad$a=a(t),\;\;f_{\pm}=f_{\pm}(t),\;\;T=\mathrm{const},\;\;M=\mathrm{const}$,}\end{center}}
with 
\be
e^\nu=[1+k/4\,(\tilde{x}^2+\tilde{y}^2)]^{-1},\,\quad k=0,\pm1,\, \quad W=1,
\ee
\be
  A = \left\{ 
  \begin{array}{l l}
    e^\nu\{a(r)[1-\frac{k}{4}(\tilde{x}^2+\tilde{y}^2)]+b(r)\tilde{x}+c(r)\tilde{y}\}-k\beta_+,\, \quad k=\pm 1\\
    a(r)+b(r)\tilde{x}+c(r)\tilde{y}-\beta_+(\tilde{x}^2+\tilde{y}^2)/2,\qquad\qquad\quad\;\,\, k=0.\\
  \end{array} \right.
\ee
The functions $a(r)$, $b(r)$, $c(r)$, and $\beta_{\pm}(r)$ are smooth but otherwise arbitrary. The function $a(r)$ is distinct from the scale factor $a(t)$.
%
\section{Relationships Between the GW and LT-like Class I Metric Representations}
%
The Szekeres metric can be written alternatively in the LT-like representation \cite{Hellaby1996}
\be
	ds^2=-\mathrm{d}t^2+\frac{(\Phi,_{r}-\Phi \mathcal{E},_{r}/\mathcal{E})^2}{\epsilon-k(r)}\mathrm{d}r^2+\frac{\Phi^2}{\mathcal{E}^2}(\mathrm{d}p^2+\mathrm{d}q^2),
\ee
where $\Phi=\Phi(t,r)$ represents an areal radius and is defined by 
\be
	(\Phi,_{t})^{2}=-k(r)+\frac{2\tilde{M}(r)}{\Phi}+\frac{\Lambda}{3}\Phi^2,\label{eq:phi}
\ee
where $\tilde{M}(r)$ corresponds to the active gravitational mass within a sphere of coordinate radius $r$, and
\be	
	\mathcal{E}(r,p,q)=\frac{1}{2}\left[\frac{(p-P(r))^2}{S(r)}+\frac{(q-Q(r))^2}{S(r)}+\epsilon S(r)\right].\label{eq:E}
\ee
The constant $\epsilon=0,\pm 1$ describes the geometry of the $(p,q)$ 2-surfaces.

If we consider a curvature of the form $k(r)=K\tilde{f}^2(r)$ ($K=|k(r)|/k(r)=0,\pm 1$) as described by \cite{Plebanski&Krasinski2006}, then we can identify $\tilde{f}(r)=\sqrt{|k(r)|}$ as the GW metric function $f(r)$. This holds except when $k(r)=0$ ($K=0$), in which case $f(r)\ne 0$ is arbitrary. We can then build a complete correspondence between the metric functions in the following two representations:

\begin{align}
a(r)&=\frac{1}{2S(r)} \nonumber\\
b(r)&=\frac{-P(r)}{2S(r)} \nonumber\\
c(r)&=\frac{-Q(r)}{2S(r)} \nonumber\\
d(r)&=\frac{P^2(r)+Q^2(r)+\epsilon S^2(r)}{2S(r)} \nonumber\\
f(r)&=\tilde{f}(r) \nonumber\\
k&=K\label{eq:correspondence}\\
\epsilon&=\epsilon \nonumber\\
a(t,r)&=\frac{\Phi(t,r)}{f(r)} \nonumber\\
M(r)&=\frac{\tilde{M}(r)}{f^3(r)} \nonumber\\
W(r)&=\frac{1}{\sqrt{\epsilon-k(r)}} \nonumber\\
H(r)&=f(r)\left[\frac{\Phi,_{r}(t,r)}{\Phi(t,r)}-\frac{\mathcal{E},_{r}(r,p,q)}{\mathcal{E}(r,p,q)}\right].\nonumber
\end{align}

The above relations have some consequences for the geometrical interpretation of the GW coordinates, following insights provided by the LT-like representation. We can immediately identify that the $(\tilde{x},\tilde{y})$ and $(p,q)$ 2-surfaces obey the same geometry in both metric representations, as indicated by $\epsilon$, and that the GW spatial curvature ($k$) in some region matches that of the LT-like spatial curvature ($K=k(r)/|k(r)|$). If we write the GW $e^{\nu}$ in a suggestive way as $e^{\nu}=f(r)/E(r,\tilde{x},\tilde{y})$, where \cite{foot1}
\be
E(r,\tilde{x},\tilde{y})=\frac{1}{2}\left[\frac{(\tilde{x}+b/a)^2}{1/2a}+\frac{(\tilde{y}+c/a)^2}{1/2a}+\epsilon\left(\frac{1}{2a}\right)\right],
\ee
we can identify the GW $\tilde{x}$ and $\tilde{y}$ as projected coordinates following the same geometry as the LT-like $p$ and $q$ (see \cite{Plebanski&Krasinski2006}, \S 19.5.3), where the GW functions $(b/a,c/a,1/2a)$ take the place of the LT-like functions $(P,Q,S)$. We can then translate the GW metric coordinates $(t,r,\tilde{x},\tilde{y})$ to both spherical $(t,r,\theta,\phi)$ and Cartesian-like $(t,X,Y,Z)$ coordinates. Within each surface of constant $t$ and $r$, these transformations are
\begin{align}
(a\tilde{x}+b,a\tilde{y}+c)&=\frac{1}{2}\cot\left(\frac{\theta}{2}\right)(\cos \phi,\sin\phi),~\quad\,(\epsilon=+1)\nonumber\\
(a\tilde{x}+b,a\tilde{y}+c)&=\frac{(\cos \phi,\sin\phi)}{\theta},\qquad\qquad\quad\;\;\;\,(\epsilon=~~0)\label{eq:convert1}\\
(a\tilde{x}+b,a\tilde{y}+c)&=\frac{1}{2}\coth\left(\frac{\theta}{2}\right)(\cos \phi,\sin\phi);\quad(\epsilon=-1)\nonumber
\end{align}
and
\begin{align}
X&=r\sin\theta\cos\phi,\nonumber\\
Y&=r\sin\theta\sin\phi,\label{eq:convert2}\\
Z&=r\cos\theta.\nonumber
\end{align}
One can see that with a simple substitution following Eq. (\ref{eq:correspondence}), Eq. (\ref{eq:convert1})  reduces directly to the transformations for the LT-like metric coordinates, given, for example, in \cite{Plebanski&Krasinski2006}.

Finally, it is clear that $a(t,r)$ and $M(r)$ play the same physical roles (albeit appropriately rescaled) as their well-studied counterparts $\Phi(t,r)$ and $\tilde{M}(r)$ in the LT-like representation. 
\begin{figure}
\begin{center}
\includegraphics[scale=0.49,angle=-0]{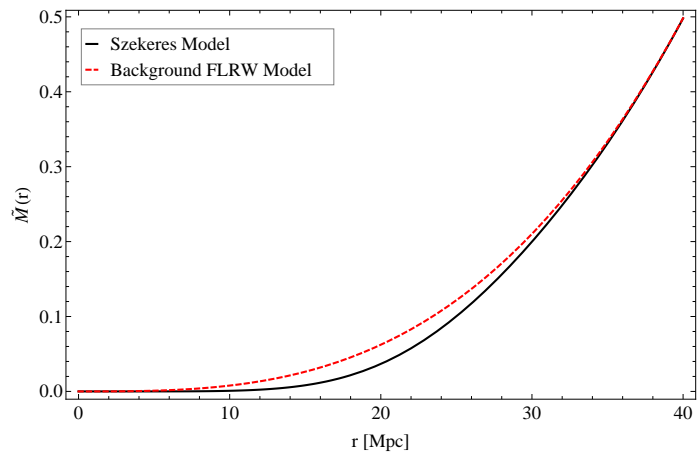}
\caption{\label{fig:m}The function $\tilde{M}(r)=f^{3}(r)M(r)$ compared to the background FLRW $\tilde{M}_b(r)=\rho_b r^3/6$. $\tilde{M}(r)$ grows less quickly than $\tilde{M}_b(r)$ at $r<15$ Mpc due to the central void, but is then compensated by an overdensity and matches to $\tilde{M}_b(r)$ at $r\ge50$ Mpc.
}
\end{center}
\end{figure}
%
\section{Modeling Exact Inhomogeneous Structures in the Class I GW Representation}
%
For studying the growth of structure in the class I GW representation of the Szekeres metric, we construct the metric functions by considering a simple inhomogeneous and nonsymmetric structure similar to that of Model 1 in \cite{Bolejko2006}. It physically represents a central void (underdensity) with a neighboring supercluster (overdensity). Constructing specific models to represent cosmic structures seems to be more intuitive in the LT-like representation due to a simpler comparison to LTB models and some previous work using this LT-like representation has been done \cite{Bolejko2007,BolejkoSussman,Bolejko&Celerier2010}. Therefore, we perform initial calculations for the model in this representation and then convert the metric functions to the GW representation using the relationships described in Appendix B. 

\begin{figure}
\begin{center}
\includegraphics[scale=0.49,angle=-0]{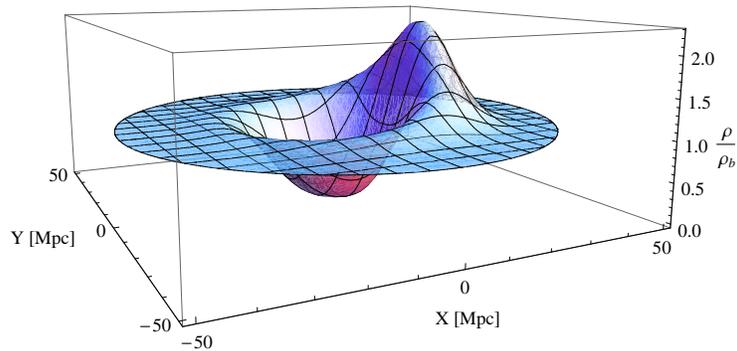}
\caption{\label{fig:mrho}The density distribution ($\rho$) given by Eq. (\ref{density}) of the model described in Appendix C is shown in units of the background density ($\rho_b$). The density is plotted in Cartesian-like coordinates for the $Z=0$ plane \cite{Zcoordinate}, converted from the metric coordinates through Eqs. (\ref{eq:convert1}) and (\ref{eq:convert2}). The model consists of a central void (underdensity) with a neighboring supercluster (overdensity).}
\end{center}
\end{figure}

Our model is matched to an almost FLRW background model described by $\Omega_m^0=0.27$, $\Omega_{\Lambda}^0=0.73$, and $\mathrm{H}_0=72$ km s$^{-1}$ Mpc$^{-1}$ at a radius of 50 Mpc. This background FLRW model has a homogeneous background density given by $\rho_b=\Omega_m \rho_{cr}$. The model is fully defined through the following series of simple steps:

\begin{enumerate}[label=(\roman*)]
 \item We begin by constructing the function
\be
	\tilde{M}(r)=f^3(r)M(r)=\frac{\rho_b}{6}r^3\left(1-\exp\left[-3\left(\frac{r}{\sigma}\right)^3\right]\right)\label{eq:Mtilde}
\ee
as shown in Fig. \ref{fig:m}, where $\sigma=30$ Mpc, such that at a radius of 50 Mpc we match to the almost FLRW background density. For comparison, $\tilde{M}_b(r)=\rho_b r^3/6$ of the FLRW background density is also shown.
 \item The metric functions P, Q, and S are chosen. In our case, we choose the functions to match those used in Model 1 of \cite{Bolejko2006}:
 \begin{align}
P(r)&=10,\nonumber\\
Q(r)&=-113\ln(1+r),\\
S(r)&=140.\nonumber
\end{align}
These define the function $\mathcal{E}(r,p,q)$ given in Eq. (\ref{eq:E}), which governs the $(p,q)$ dependence of the resulting density distribution of the structure, and $\epsilon=+1$ for this model.
 \item The $r$ coordinate is defined such that $\Phi(t_0,r)=r$, where $t_0$ is the age of the universe today.
 \item We integrate Eq. (\ref{eq:phi}) in order to find $k(r)$,
\be
	\int^r_0\frac{\mathrm{d}\Phi}{\sqrt{-k(r)+\frac{2\tilde{M}(r)}{\Phi}+\frac{\Lambda}{3}\Phi^2}}=t_0,\label{eq:k}
\ee
where we have chosen the bang time to be everywhere zero (i.e., $t_B(r)=0$). 
\item The function $\Phi(t,r)$ is now fully determined by Eq. (\ref{eq:phi}).
\item Now that $\tilde{M}(r)$, $k(r)$, and $\Phi(t,r)$ are known for the model, we can freely transform to the GW metric representation following the relations given in our Appendix B.
\end{enumerate}
The density distribution of the model is given by Eq. (\ref{density}), where $\hat{\delta}(t,r,\tilde{x},\tilde{y})$ is evaluated as discussed in Sec. \ref{growthEqnsClassI}. The resulting density distribution in the $Z=0$ plane \cite{Zcoordinate} is shown in terms of the background density in Fig. \ref{fig:mrho}. Though the model is designed to match a $\Lambda$CDM background with zero curvature, the Szekeres region itself has a nonzero curvature.
%

%

\begin{thebibliography}{}
\bibitem{foot1} As in Appendix A, the function $a$ used here is $a(r)$ and is distinct from the generalized scale factor $a(t,r)$.
\end{thebibliography}
\end{document}